\documentclass[paper]{JHEP3}

\usepackage{epsfig,multicol,bbm}

\newcommand\fverb{\setbox\fverbbox=\hbox\bgroup\verb}
\newcommand\fverbdo{\egroup\medskip\noindent%
            \fbox{\unhbox\fverbbox}\ }
\newcommand\fverbit{\egroup\item[\fbox{\unhbox\fverbbox}]}
\newbox\fverbbox

\title{Curing singularities in cosmological evolution of $F(R)$ gravity}

\author{$^{1}$Stephen A. Appleby, $^{1}$Richard A. Battye,
$^{2}$Alexei A. Starobinsky\\
$^{1}$Jodrell Bank Centre for Astrophysics, School of Physics and Astronomy \\
University of Manchester, Oxford Road, Manchester, UK, M13 9PL\\
$^{2}$L. D. Landau Institute for Theoretical Physics, Moscow
119334, Russia \\   E-mail: \email{sappleby@jb.man.ac.uk},
\email{rbattye@jb.man.ac.uk}, \email{alstar@landau.ac.ru}}

\abstract{We study $F(R)$ modified gravity models which are
capable of driving the accelerating epoch of the Universe at the
present time whilst not destroying the standard Big Bang and
inflationary cosmology. Recent studies have shown that a weak
curvature singularity with $|R|\to\infty$ can arise generically in
viable $F(R)$ models of present dark energy (DE) signaling an
internal incompleteness of these models.
In this work we study how this problem is cured by adding a
quadratic correction with a sufficiently small coefficient to the
$F(R)$ function at large curvatures.
At the same time, this correction eliminates two more serious
problems of previously constructed viable $F(R)$ DE models:
unboundedness of the mass of a scalar particle (scalaron) arising
in $F(R)$ gravity and the scalaron overabundance problem. Such
carefully constructed models can also yield both an early time
inflationary epoch and a late time de Sitter phase with vastly
different values of $R$. The reheating epoch in these combined
models of primordial and present dark energy is completely
different from that of the old $R + R^{2}/6M^{2}$ inflationary
model, mainly due to the fact that values of the effective
gravitational constant at low and intermediate curvatures are
different for positive and negative $R$. This changes the number
of e-folds during the observable part of inflation that results in
a different value of the primordial power spectrum index.}

\begin{document}

\section{\label{sec:1}Introduction}

Numerous recent observational data prove convincingly that the
Universe is undergoing accelerated expansion at the present time, whilst
 decelerating in the past for redshifts larger than about
$z \sim 0.7$. If interpreted in terms of the Einstein general theory of
relativity, this acceleration requires the existence of some new
component in the right-hand side of the Einstein equations, dubbed
dark energy (DE), which remains practically non-clustered at all
scales at which gravitational clustering of baryonic and dark
non-baryonic matter is observed, and which has an effective pressure $p_{DE}$
 approximately equal to minus its effective energy density
$\rho_{DE}$. Thus, its properties are very close to those of a
cosmological constant $\Lambda$ (see
\cite{SS00,P02,CST06,SS06,FTH08} for some reviews). The simplest
possible DE model, $\Lambda$ combined with a
non-relativistic non-baryonic dark matter (the standard spatially
flat $\Lambda$CDM cosmological model), is completely
self-consistent from the mathematical point of view and provides a
good fit to all existing observational data \cite{K08}. In this
case $\Lambda$ acquires the status of a new fundamental physical
constant. However, its required value is very small compared to
known atomic and elementary particle scales (which are well below the Planck scale), so a firm theoretical prediction for this
quantity from first principles is currently lacking.

On the other hand, in the second case when a component with
qualitatively similar properties is assumed to exist -- in the
inflationary scenario of the early Universe, we are sure that this
``primordial DE" may not be an exact cosmological constant since
it should decay in the early Universe. Hence it is natural to seek
non-stationary models of the current DE, too.

Among them, the simplest purely gravitational models in $3+1$
space-time dimensions are provided by $F(R)$ gravity which
modifies and generalizes Einstein gravity by incorporating a
new phenomenological function of the Ricci scalar $R$, $F(R)$.
They represent a self-consistent and non-trivial alternative to the
$\Lambda$CDM model. The literature on these models is dense, and we
direct the reader to \cite{so1} and references therein for a
detailed recent review. The action of $F(R)$ gravity is given by
\begin{equation}
S=\frac{M_{\rm Pl}^2}{2}\int d^4x \sqrt{-g}F(R) +S_m~, \label {S}
\end{equation}
where $M_{\rm Pl}^2=1/8\pi G$, $\hbar =c=1$ is assumed throughout
the paper, and $S_m$ describes all non-gravitational
matter including non-relativistic (cold) dark matter which is
minimally coupled to gravity.\footnote {The sign conventions are:
the metric signature (-+++), the curvature tensor $R^{\sigma}_{\
\mu\rho\nu} = \partial_{\rho}\Gamma^{\sigma}_{\mu\nu} - ...,
~R_{\mu\nu}=R^{\sigma}_{\mu\sigma\nu}$, so that the Ricci scalar
$R=R_{\mu}^{\mu}>0$ for the de Sitter space-time and the
matter-dominated cosmological epoch.} The field equations
following from (\ref{S}) have the form
\begin{equation}
F'(R)R_{\mu}^{\nu} - \frac{1}{2}F(R)\delta_{\mu}^{\nu}+
(\delta_{\mu}^{\nu}\Box - \nabla_{\mu}\nabla^{\nu})F'(R)= M_{\rm
Pl}^{-2}T_{\mu}^{\nu}~, \label{FReq}
\end{equation}
and their trace reads
\begin{equation}
3\Box F'(R) + RF'(R) - 2F(R)= M_{\rm Pl}^{-2}T~, \label{trace-eq}
\end{equation}
where $T$ is the trace of the matter energy-momentum tensor
$T_{\mu}^{\nu}$ and the prime denotes the ordinary derivative with
respect to an argument.

In the Jordan frame where the action (\ref{S}) is written, fermion
masses are constant and atomic clocks built from usual matter
measure the proper time. However, equivalent description is
possible in the Einstein frame where gravity resembles standard GR
but free particles of usual matter do not follow space-time
geodesics due to an interaction with a new scalar field. As
will be seen below, sometimes it is easier to solve equations in
the latter frame first. We also assume the metric variation of
(\ref{S}); the Palatini variation of formally the same action
leads to a completely different theory, in which the number of degrees
of freedom is not the same.

In the absence of matter, exact de Sitter (positive constant
curvature) solutions of Eqs. (\ref{FReq}) are given by real
positive roots of the functional equation
\begin{equation}
RF'(R)-2F(R)=0~. \label{dS}
\end{equation}
These solutions (and solutions close to them) are the basis for a
description of primordial and present DE. They are future stable
if
\begin{equation}
F'(R_1)/F''(R_1)>R_1~, \label{dSstab}
\end{equation}
where $R_1$ is a root of Eq. (\ref{dS}). This condition was first
obtained in \cite{MSS88}, the easiest way to derive it is to use the
trace equation (\ref{trace-eq}) for a small perturbation $R-R_1$
(in fact, $F''(R_1)$ should be positive, too, as will be discussed
below).

$F(R)$ gravity is a special class of scalar-tensor gravity with
a vanishing Brans-Dicke parameter $\omega_{BD}$. If $F''(R)$ is
not zero identically, it contains a new scalar degree of freedom
dubbed ``scalaron" in \cite{st1}, thus, it is a {\em
non-perturbative} generalization of Einstein gravity. We will
consider it as a purely phenomenological semiclassical macroscopic
theory of gravity which arises from some more fundamental quantum
microscopic theory after tracing out degrees of freedom which are
not excited at sufficiently small space-time curvature. Thus, the
resulting function $F(R)$ need not necessarily be some simple
(e.g. polynomial) function of $R$. It may well have some
complicated behaviour for small $R$, too, as many examples from
condensed matter physics teach us. So, we will not discuss which
functional form of $F(R)$ is ``natural" in any sense. On the other
hand, for a phenomenological $F(R)$ model to be
viable, it should satisfy a rather large list of viability
conditions:

1) Classical and quantum stability in the region of $R$ where we
want to use this theory:
\begin{equation}
F'(R)>0,~F''(R)>0~. \label{stabil}
\end{equation}
The first condition means that gravity is attractive and the graviton
is not a ghost. It was recognized long ago that its violation
during the time evolution of a Friedmann-Robertson-Walker (FRW)
background results in the immediate loss of homogeneity and
isotropy and formation of a strong space-like anisotropic
curvature singularity \cite{H73,GS79}. The second condition on the
flat background was also known since the first papers on $F(R)$
gravity \cite{RR70}, and was assumed when constructing
inflationary models in $F(R)$ gravity \cite{st1}. However, in
the case of $F(R)$ models of present DE, the necessity to keep it
valid for all values of $R$ during the matter- and
radiation-dominated stages in order to avoid the Dolgov-Kawasaki
instability \cite{d1} has been realized rather recently
\cite{at10,am1}. In addition, a weak (``sudden") curvature
singularity forms generically if $F''(R)$ becomes zero for a
finite value $R=R_s$. This is also undesirable; see the discussion
below in Sec. 2.4.

2) Existence of the stable Newtonian limit for all values of $R$
where Newtonian gravity accurately describes observed
inhomogeneities and compact objects in the Universe, i.e. for $R$
exceeding the present Friedmann-Robertson-Walker (FRW) background
value $R_0\equiv R(t_0)$, where $t_0$ is the present moment, and
up to curvatures in the centre of neutron stars. The conditions
required for this are
\begin{equation}
|F(R)-R|\ll R~,~~|F'(R)-1|\ll 1~,~~RF''(R)\ll 1 \label{GR}
\end{equation}
for $R\gg R_0$. Note that in this regime the effective scalaron
mass squared is $M_s^2(R)=1/(3F''(R))$, as directly follows from
Eq. (\ref{trace-eq}). Then the second of the conditions
(\ref{stabil}) means that the scalaron is not a tachyon, while the
last of the conditions (\ref{GR}) implies that its Compton
wavelength is much less than the radius of curvature of the
background space-time. For more general backgrounds than
matter-dominated FRW, for which General Relativity (GR) has to
be used in full (in particular, if the pressure $P$ of matter is not
small compared to its energy density $\rho$), the conditions
(\ref{GR}) guarantee that non-GR corrections to a
space-time metric remain small.

3) Absence of deviations from GR at the level of accuracy
following from present laboratory and Solar system tests of
gravity.

4) Existence of a future stable (or at least metastable) de Sitter
asymptote. This is necessary for a description of the present DE,
which behaves in a similar manner to that of a cosmological constant.

5) $F(R)$ cosmology should not destroy previous successes of
present and early Universe cosmology in the scope of GR including
the existence of the matter-dominated stage driven by
non-relativistic matter preceded by the radiation-dominated stage
with the correct Big Bang nucleosynthesis (BBN) of light elements
and, as we shall see, some kind of inflation prior to this.

The first of these conditions is also the main reason why we
do not include other invariants constructed from the Riemann
tensor and its derivatives as additional arguments of the function
$F$. Indeed, it has long been known
\cite{UD62,Stel77} that if $F$ also depends on
$R_{\mu\nu\alpha\beta}R^{\mu\nu\alpha\beta}$, then
generically a new massive spin-2 particle appears which is a ghost
whenever the standard massless graviton is not a ghost. This is problematic from the quantum
field theory point of view for many reasons, see e.g.
\cite{CJM04}. Moreover, as argued in \cite{GPRS06}, cosmological
models with ghosts are unsatisfactory even at the purely classical
level. In particular, in such models we can no longer
explain the observed approximate large-scale homogeneity and
isotropy of the Universe without tremendous fine-tuning of initial
conditions, even if we include a primordial inflationary stage. The only
way to avoid this ghost (without including derivatives of the Riemann tensor in the action) is to consider $F=F(R,G)$ where
$G$ is the Gauss-Bonnet invariant,
$G=R_{\mu\nu\alpha\beta}R^{\mu\nu\alpha\beta}-4R_{\mu\nu}R^{\mu\nu}+R^2$.
However, it was recently shown in \cite{FS09} that in this case linear
scalar perturbations have pathological behaviour in the
ultra-violet regime, except in some special cases.\footnote{After
the first variant of our paper was submitted to JCAP and archives,
the further paper \cite{FMT10} appeared where this remaining
special case was shown to possess an ultra-violet instability of
scalar perturbations on a FRW background with matter in the form
of a perfect fluid, too.} Invariants in $F$ containing derivatives
of the Riemann tensor lead to new particles (in particular, scalar
ones if $F$ depends on $R$ and its derivatives only \cite{GSS90})
among which ghosts are generic, too, see e.g. \cite{HOW96}. Thus,
such terms may be considered when making perturbative expansion
around solutions of the Einstein gravity, but (possibly apart from
some exceptional cases still to be found) they are of no use in
our approach since we want to use a (semi)classical modified
theory of gravity in a fully non-perturbative regime.

 For these reasons, we concentrate on $F(R)$ effective
macroscopic gravity only. Note that this is in contrast to
quantum-gravitational and string corrections to Einstein
gravity, which generically produce terms with all possible
invariants of the Riemann tensor. The required
``$R$-dominance" presents a serious problem for such microscopic
mechanisms to act as the origin of $F(R)$ gravity, however there exists a number
of cases when just this form of modified gravity appears in some
limit. The simplest of them follows from above-mentioned fact that
$F(R)$ gravity is a particular case of more general scalar-tensor
gravity with the Brans-Dicke parameter $\omega_{BD}=0$. Therefore,
it also yields a good approximate description for scalar-tensor
gravity with $|\omega_{BD}|\ll 1$, in particular, for a
non-minimally coupled scalar field with a negative and large by
modulus coupling constant $\xi$ (we use the sign convention where
conformal coupling corresponds to $\xi = +1/6$), cf. \cite{DSS08}.
For this reason, in particular, predictions of the Higgs
inflationary model \cite{BS08} (without loop corrections to the
Higgs potential) for primordial power spectra of scalar (density)
perturbations and gravitational waves are the same as those of the
$F(R)=R+R^2/6M^2$ model -- the simplified variant of the model
\cite{st1}. Another, completely unrelated case where
phenomenological $F(R)$ gravity arises rather unexpectedly
\cite{KV08} is the so called emergent gravity approach using ideas
and methods borrowed from quantum theory of condensed matter. This
illustrates the well-known fact that an elegant and internally
consistent mathematical model may appear multiple times from
totally different physical foundations. Note that in both of these
examples, the new scalar gravitational degree of freedom
(scalaron) is present already at the underlying microscopic level;
however, this does not mean that it is fundamental even at this
level.

Although $F(R)$ gravity can successfully pass the first requirement
from the list above, it is evident from the beginning that only a
very narrow subset of all possible $F(R)$ functions may be of
interest for cosmology. The situation is simpler in case of
inflationary models. Here, the simplest variant with $F''(R)\not=
0$ identically, namely $F=R+R^2/6M^2$ (where $M$ is the scalaron
rest-mass at low curvature), presents an internally
self-consistent inflationary model with slow-roll during
inflation (see Sec.~4.4 below) and a graceful exit to a subsequent
FRW matter-dominated stage driven by scalarons \cite{st1}.
Reheating, creation of usual matter and transition to the
radiation-dominated FRW stage are achieved by gravitational
particle production due to strong oscillations of $R$ during the
scalaron-dominated stage \cite{st1}, see
\cite{S82,vk1,mj1,mj2,FTBM07} for more details regarding the background
evolution and reheating in this model. In addition, in contrast to
many other inflationary models proposed later which have been
falsified by observational data, this model still remains viable
since it predicts the value of the slope of the primordial power
spectrum of scalar (density) perturbations $n_s=1 - 2/N$
\cite{MC81,st3} in agreement with present observational data. Here
$N$ is the number of e-folds between the first Hubble radius
crossing of the present inverse comoving scale $0.002$ Mpc$^{-1}$
and the end of inflation; $N\approx (50-55)$ for the reheating
mechanism mentioned above. The tensor-to-scalar ratio is rather
small, $r=12/N^2$, but not negligible \cite{st3}; see also
\cite{st2,KMP87,HN01} for further papers on equations and
solutions for perturbations and \cite{se1} for the energy theorem
in this model. To fit the observed amplitude of the power
spectrum, the only free model parameter $M$ should be chosen as
$M\approx 1.5\times 10^{-5}(N/50)^{-1}M_{\rm Pl}$. Furthermore, we
will show in Sec. 4.4. that all viable inflationary models in
$F(R)$ gravity with other values of $n_s$ and $r$ have behaviour
close to $R^2$ at large $R$ (or around some large fixed value
$R=R_{1}$). Thus, the $R^2$ behaviour of $F(R)$ is characteristic
for inflation in $F(R)$ gravity.

Due to the remarkable quantitative analogy between properties of
primordial DE supporting inflation in the early Universe and
present DE, it is tempting to use $F(R)$ gravity to also build models
of the latter, to act as alternatives to the trivial case of a
cosmological constant which corresponds to $F(R)=R-2\Lambda$.
Indeed, many such models have been proposed, beginning with the
papers \cite{C02,CCT03,CDTT04,NO03}. However, most of these
attempts remained unsuccessful since the conditions written above
had not been fully satisfied (once more, we direct a reader to the
review \cite{so1} for an extensive list of papers on the subject).
As a result, there has arisen a widespread doubt if viable $F(R)$
models of present DE may exist at all (in contrast to primordial
DE).

Still at last a rather narrow class of functional forms of $F(R)$
was found \cite{hu,ap,st} which can satisfy the first four
viability conditions from the list above and even partly the fifth
one, with regards to the existence of a stable matter
dominated epoch with $a(t) \propto t^{2/3}$ in the recent past
which is driven by cold dark matter and baryons (not, we stress, by terms due
to modified gravity.) Here, $a(t)$ is the scale factor in a FRW
spacetime. It should be emphasized from the very beginning that in
none of these viable $F(R)$ DE models is it possible to derive the
energy scale of present DE from first principles. Instead, it has
to be inserted into the action (\ref {S}) as a free parameter, the value of which is taken from observational data. Thus, $F(R)$ DE
models may not be superior to DE being an exact cosmological
constant $\Lambda$, they are simply an alternative to it. In the
models \cite{hu,ap,st}, $F(R)$ is analytic for $R=0$, has a
non-trivial structure for $R\sim R(t_0)$ and then quickly
approaches Einstein gravity with an effective cosmological
constant for $R\gg R_0$, see Eq. (\ref{largeR}) below. Also, in
these models the condition $F(0)=0$ (dubbed the ``disappearing
cosmological constant" in \cite{st}) is imposed by hand to ensure that there is no true cosmological constant in flat
space-time (otherwise, why work with $F(R)$ gravity at all?) The
main difference between these models is in the law of approach to
the standard gravity for $R\to\infty$: an inverse power law in
\cite{hu,st} and an exponential in \cite{ap}.

However, it was immediately recognized that the story of
constructing at least one viable cosmological model of present
DE in $F(R)$ gravity was not finished: three new problems related
to the last, fifth viability condition arise when tracing small
deviations of a FRW background from the standard $\Lambda$CDM
model to the past. First, as was shown in \cite{st} (see also
\cite{ts10}), for these models the frequency of small oscillations
of the Ricci scalar $R$ (i.e. the scalaron rest-mass) around the
general relativistic limit $R_{\rm GR} = -T / M_{\rm Pl}^{2}$,
$\omega\equiv M_s(R)=1/\sqrt{3F''(R)}$, grows quickly to the past,
$t\to 0$, and exceeds the Planck value very soon, thus
invalidating classical consideration of the theory (\ref{S}). For
example, for the model \cite{ap} it happens already for redshifts
$z>7$ and matter densities exceeding $\sim 10^{-27}$ g cm$^{-3}$.

The second problem is that the amplitude of these linear
oscillations quickly grows back in time \cite{st}. As a
result, we get the scalaron overabundance problem which is
actually a problem of initial conditions of the Universe; they
should be such that the scalaron number density (which is proportional to
the square of an amplitude of these oscillations) should be
sufficiently small at the period of BBN. Third, due to the same
reason, linear consideration of these oscillations may become
inadequate even before their backreaction on a FRW background
becomes important. A non-linear analysis of the these models was
undertaken in \cite{ap2}, where it was observed that the Ricci
scalar would generically evolve to a weak singularity at some
finite time in the past. This singular behaviour was predicted
independently in \cite{afq1} for a more general class of models.
These three problems once again raised the question as to whether viable DE $F(R)$ gravity models can be constructed.

The main aim of this paper is to find a way to avoid these three
difficulties and to show that there indeed exist $F(R)$ DE models
satisfying all five viability conditions. We shall show that for
this purpose, as envisaged in \cite{st}, it is sufficient to
change the behaviour of $F(R)$ at large $R$ by adding the term
$\propto R^2$ with a sufficiently small coefficient to ensure the existence of an inflationary stage in the
early Universe. Moreover, since this behaviour is
characteristic for viable inflationary models in $F(R)$ gravity,
we show additionally that it is possible to construct a combined
model where both primordial and present DE are described in the
scope of $F(R)$ gravity.\footnote{Of course, this model does not
really unify primordial and present DE since we have to introduce
two tremendously different curvature scales corresponding to
inflation in the early Universe and to the present space-time
curvature by hand.} Rather unexpectedly, such a construction also
requires us to change the low curvature behaviour of $F(R)$ in the
models \cite{hu,ap,st} for $R<R_0$ and further to the region of
negative values of $R$ which are not observable at the present
time, in order to avoid violation of the conditions (\ref{stabil})
during strong oscillations of $R$ after the end of inflation.
Moreover, it will be shown that a non-trivial structure of $F(R)$
at low $R$, required for an alternative description of present DE (as opposed to a
cosmological constant), greatly affects the stage of
post-inflationary evolution and reheating in this combined model,
and even results in the change of numerical values for parameters
of primordial power spectra of perturbations generated after
inflation.

The rest of the paper proceeds as follows. In section \ref{sec:2} we
review the cosmological evolution of viable models of present DE
in $F(R)$ gravity, and comment on the existence and properties of
the weak singularity mentioned above. We then consider possible
approaches to eliminate this singularity and bound the scalaron
mass by introducing additional terms into the action in section
\ref{sec:3}. Finally, we consider the possibility that these
additional terms may drive an early inflationary period of the
Universe, and study the slow-roll and reheating epochs in section
\ref{sec:4}. Section \ref{sec:5} contains conclusions and
discussion.

\section{\label{sec:2}Review of cosmological evolution}

We will be concerned with the following $F(R)$ functions
describing present DE which satisfy the first four viability
conditions listed above and possess a stable matter-dominated
epoch at intermediate redshifts $z$ for some range of their
parameters:
\begin{eqnarray}\label{eq:cr1} & & F_{\rm HSS}(R) = R - {R_{\rm vac}
\over 2}{c \left({R \over R_{\rm vac}}\right)^{2n} \over 1+
c\left({R \over R_{\rm vac}}\right)^{2n}} ,\\ \label{eq:cr201} & &
F_{\rm AB}(R) = {R \over 2} + {\epsilon_{\rm AB} \over 2}
\log\left[{\cosh\left({R \over \epsilon_{\rm AB}}-b\right) \over
\cosh b}\right] ,\end{eqnarray}

\noindent where $b$ and $c$ are dimensionless constants,
$\epsilon_{\rm AB} = R_{\rm vac}/(b+\log (2\cosh b))$ and $n>0$.
The model (\ref{eq:cr1}) is the model introduced in \cite{hu}
(with a slightly different notation of parameters); it is similar
to the model of \cite{st}. The model (\ref{eq:cr201}) is from the
paper \cite{ap} (a model with a similar behaviour was also
introduced later in \cite{ts10}). For large $R$, these models
mimic General Relativity with a cosmological constant, in the
sense that for $R \gg R_{\rm vac}$, $F_{\rm HSS}$ and $F_{\rm AB}$
can be expanded in the region of interest to cosmology as

\begin{equation}\label{largeR} F(R) \approx R - {R_{\rm vac} \over 2}
+ \chi(R)~ . \end{equation}

\noindent Thus, $R_{\rm vac}/4$ acts as a small effective
cosmological constant induced by space-time curvature. Since
$F(0)=0$, there is no true cosmological constant in these models.
$\chi(R)$, $\chi'(R)$ and $\chi''(R)$ are all small functions of
$R$, which satisfy $\chi(R)/R \ll 1$, $\chi'(R)\ll 1$ and
$R\chi''(R) \ll 1$ for $R \gg R_{\rm vac}$. For the HSS and AB
models, we have

\begin{eqnarray}\label{eq:cr50} \chi_{\rm HSS} = {\epsilon^{2n+1}_
{\rm HSS}\over R^{2n}}, \qquad    \chi_{\rm AB} = {\epsilon_{\rm
AB} \over 2} e^{2b} \exp\left(-2 R/ \epsilon_{\rm AB}\right),
\end{eqnarray}

\noindent where $\epsilon_{\rm AB}$ and $\epsilon_{\rm HSS} =
R_{\rm vac}/(2c)^{1/(2n+1)}$ are smaller than $R_{\rm vac}$.

In this section we briefly review the cosmological evolution of
the AB and HSS models, in particular the behaviour of the scalar
degree of freedom (scalaron), and highlight the existence of a
singularity in the evolution of the Ricci scalar. In
ref.\cite{st}, the trace equation ($\ref{trace-eq}$) was solved
using a perturbative approach: an ansatz $R = R_{\rm GR} + \delta
R$ was taken, where $R_{\rm GR} = -T / M_{\rm Pl}^{2}$, and the
equation linearized for $\delta R \ll R_{\rm GR}$. It was found
that the Ricci scalar for the HSS model will generically undergo
rapid oscillations around its General Relativistic limit, a result
also obtained for the AB model in ref.\cite{ts10} (although see
ref.\cite{ap2} for a discussion of the applicability of the
linearized approach in the case of the AB model.) The frequency
and amplitude of these oscillations increase without bound for
both models as $R$ grows to the past ($t\to 0$). Specifically,
$\delta R$ contains an oscillating component that is given by

\begin{equation} \label{R-osc} \delta R_{\rm osc} = C a^{-3/2}
(F''(R_{\rm GR}))^{-3/4} \sin \left[ \int {dt \over \sqrt{3
F''(R_{\rm GR})}} \right] .
\end{equation}

\noindent It represents scalaron oscillations (particles) with the
frequency (rest-mass) $M_s(R)= (3F''(R_{\rm GR}))^{-1/2}$ in the
regime when $M_s^2\gg |R|_{GR}$. Typically, $M_s$ becomes $\gg
M_{\rm Pl}$ during the matter and radiation eras for both models.

\subsection{Non-linear oscillations and existence of a ``sudden"
singularity with $|R|\to\infty$}

In refs.\cite{ap2}, an alternative approach to solving
($\ref{trace-eq}$) was considered. In this work it was noted that
at curvatures of cosmological interest, we can use the expansions
($\ref{eq:cr50}$) to write ($\ref{trace-eq}$) as a non-linear
oscillator equation. By making the field redefinitions

\begin{eqnarray}\label{eq:o1} R &=& R_{\rm GR} - {\epsilon \over 2} \log (1+x) ,
\\  \label{eq:o2} R &=& {R_{\rm GR} \over (1+x)^{1/(2n+1)}},  \end{eqnarray}

\noindent for the AB and HSS models respectively, and neglecting
terms of order ${\cal O}(\chi(R_{\rm GR}))$, ($\ref{trace-eq}$)
can be written as

\begin{equation}\label{eq:l10}  x_{\lambda\lambda} + {\alpha(\lambda) \over \lambda}
x_{\lambda} + {\epsilon \over 6} \log (1+x) \approx 0  \qquad ({\rm AB}),
\end{equation}

\begin{equation} \label{eq:hu127}  x_{\lambda\lambda} + {\beta(\lambda) \over \lambda}
x_{\lambda}  - {\epsilon \over 6} \left({1 \over (1+x)^{1/3}}-1 \right)  \approx 0
\qquad ({\rm HSS}),\end{equation}

\noindent for $R \gg R_{\rm vac}$, where we have taken $n=1$ for
simplicity in the HSS model. In (\ref{eq:l10}) and
(\ref{eq:hu127})~$x$ is a dimensionless field, $\epsilon$ has the
same dimensions as the Ricci scalar, and $\alpha(\lambda)$ and
$\beta(\lambda)$ are dimensionless damping terms that will be
unimportant in the following discussion. $\lambda$ subscripts
denote derivatives with respect to the `fast time' $\lambda$,
which is related to the cosmological time by

\begin{equation} {d \lambda \over d t} = {1 \over \sqrt{|f'_{0}|}} ,
\end{equation}

\noindent where we have introduced the dimensionless function
$f'_{0} = F'(R_{\rm GR}) - 1$. Both ($\ref{eq:l10}$) and ($\ref{eq:hu127}$)
are damped, non-linear oscillator equations, with potentials given by
integrating the expressions $dV/dx = (\epsilon/6) \log(1+x)$ and
$dV/dx = (\epsilon/6) (1 - (1+x)^{-1/3})$ for the AB and HSS models
respectively.

Equations ($\ref{eq:l10}$) and ($\ref{eq:hu127}$) can be solved
numerically, and $R$ obtained from ($\ref{eq:o1}$) and
($\ref{eq:o2}$). By taking into account non-linear terms in
($\ref{trace-eq}$), a number of effects were observed that were
not apparent in the linearized analysis. Specifically, it was
found the oscillations become asymmetric in the past (as expected,
when the linearized approximation breaks down), and further that
by evolving $R$ backwards in time, it will generically evolve to a
singularity.

A singularity arises in both models as $x \to -1$ after a finite
time. At this point, the potential has the asymptotic behaviour
$V(x) \to {\rm const}$, but its derivative diverges $dV/dx \to
\infty$. From ($\ref{eq:o1}$) and ($\ref{eq:o2}$), it is clear
that the Ricci scalar will diverge as $x \to -1$. The potentials
for the AB and HSS models are given by

\begin{eqnarray}\label{eq:kd1} V_{\rm AB}(x) &=& \alpha_{1} +
{\epsilon \over 6} (1+x) \left[\log(1+x) - 1 \right] ,\\ \label{eq:kd2}
V_{\rm HSS}(x) &=& \alpha_{2}+ {\epsilon \over 6} \left[(1+x) -
{2n+1 \over 2n}(1+x)^{2n/(2n+1)} \right] ,\end{eqnarray}

\noindent and are exhibited in fig.\ref{fig:1}. $\alpha_{1,2}$ are
integration constants, which dictate the ground state of $x$. In the
limit $x \to -1$, we have $V \to {\rm const}$, as expected.

\FIGURE{\hspace{10mm}\epsfig{file=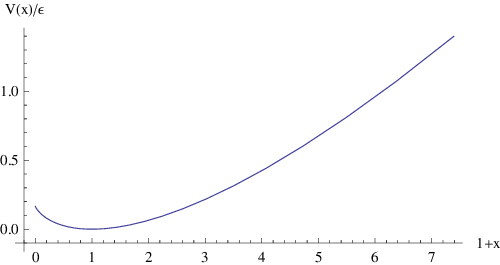,width=0.7\textwidth}\\
        \epsfig{file=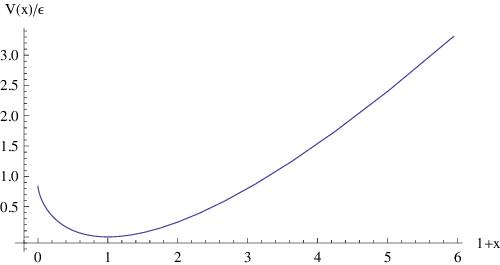,width=0.7\textwidth}
        \caption[]{\label{fig:1}The potential ($\ref{eq:kd1}$) for
        the AB model (top), with $b = 4$, $R_{\rm GR} = 10 R_{\rm vac}$
        and $\alpha_{1} = 0$, and HSS model ($\ref{eq:kd2}$) (bottom),
        with $n=2$, $c=1$, $R_{\rm GR} = 10 R_{\rm vac}$ and $\alpha_{2}
        = 0$. The singular point is at $x \to -1$, where the potential
        is finite but $dV/dx$ diverges.  }
    }

The divergence of the Ricci scalar represents a weak singularity
in the sense that $\ddot{a}$ and $R$ diverge whilst $\rho,~p$ and
$\dot{a}$ remain finite. As a result, a small vicinity of
space-time containing this space-like singularity (and from both
sides of it) may be covered by the Minkowski metric with small
perturbations which, however, are not $C^2$ continuous. Moreover,
if the singularity occurs during the matter-dominated epoch when
$p\ll \rho$, the Newtonian approximation is applicable around
it. Let us obtain the expression for the behaviour of a FRW scale
factor near the singularity. It follows from the trace equation
Eq. (\ref{trace-eq}) that the analytic form of $a(t)$ in the
vicinity of $t=t_{\rm s}$ -- the time at which $R$ diverges --
is given by

\begin{eqnarray} a(t) &=& a_{0} + a_{1} (t-t_{\rm s}) + a_{2} (t-t_{\rm
s})^{2}\left(\log |t-t_{s}|+ \tilde a_2 \right) + ...~,\label{BB1}\\
a(t) &=& a_{0} + a_{1}(t-t_{\rm s}) + a_{2} |t-t_{\rm
s}|^{(1+4n)/(1+2n)} + ...~,\label{BB2}\end{eqnarray}

\noindent for the AB and HSS models respectively, where
$a_{0,1,2}$ and $\tilde a_2$ are constants. Such a singularity was
considered in \cite{B04} from a kinematic viewpoint (i.e. not as a
solution of any dynamical equations) where it was called ``sudden".
It also appeared in a different dynamical setting in \cite{BDK08},
where it was dubbed the ``Big Boost".

In contrast to strong curvature singularities occurring in
cosmology and inside black holes, there is no geodesic
incompleteness here \cite{jam1}. However, this does not mean that
the weak ``sudden" singularities (\ref{BB1},~\ref{BB2}) are
harmless. Just the opposite, they are undesirable. Indeed, it can
be checked that Eqs. (\ref{FReq}) do not supply us with any
information on how the coefficients $a_2$ for $t<t_{\rm s}$ and
$t>t_{\rm s}$ are related to each other. Thus, the "sudden"
singularity results in the loss of predictability; from initial
data given at any Cauchy hypersurface with $t=const<t_{\rm s}$, it
is not possible to predict the space-time metric for $t>t_{\rm s}$
unambiguously. Note also that in contrast to strong curvature
singularities, we may not evade the problem by arguing that
quantum-gravitational effects invalidate the very notion of
deterministic classical space-time near such a singularity. It can
be shown \cite{bar1} that such effects remain small as one approaches the
weak singularity; the energy density and other components of an
averaged energy-momentum tensor (EMT) of quantum
fields in such curved space-time generically remain subdominant to
the background (this also follows from general expressions for an
EMT average value in a weakly curved background obtained in
\cite{stz}). So, the singularity has to be resolved at the
classical level.

Therefore, the appearance of ``sudden" singularities in a given
$F(R)$ DE model signals the internal incompleteness of said model.
One way to resolve the singularity is to return from the effective
macroscopic $F(R)$ theory (\ref{S}) to the underlying microscopic
theory from which the former originated, and see what happens
in the latter. Then, however, there is no guarantee that the
resulting ansatz for the resolution of the ``sudden" singularity
will be universal and will not depend on the choice of underlying
physics. Since we prefer to remain at the purely phenomenological
level in this paper, we instead look for a way to modify the given
$F(R)$ model in such a way that the weak singularities do not
appear at all, at least in solutions which are of interest for
cosmology, an approach that we will consider in Sec. 3.

\subsection{\label{sec:nf1}General viable $F(R)$ models of present DE}

Thus far, we have focussed attention on two specific models. We
will now generalize the existence of a singularity to an arbitrary
$F(R)$ function for which $F''(R)>0$ and $R|F'(R)-1| \to 0$ as $R
\to \infty$. Whilst viable $F(R)$ models do not have to satisfy
this asymptotic limit at infinity, they do have to satisfy
$RF''(R) \ll 1$ for $R \gg R_{\rm vac}$, as we demand that any
modification to General Relativity should only become dominant at
late times. We begin with the function $F(R) = R - R_{\rm vac} / 2
+ \chi(R)$, where $\chi(R)$ satisfies $\chi(R)/R \ll 1$, $\chi'(R)
\ll 1$ and $R\chi''(R) \ll 1$. We then define $1+x =
\chi'(R)/\chi'_{0}$, where $\chi'_{0} \equiv \chi'(R_{\rm GR})$,
and write ($\ref{trace-eq}$) in terms of $x$,

\begin{equation}\label{eq:i2} 3\Box  \chi'_{0}(1+x) - (R - R_{\rm GR})
+ \chi'_{0}R (1+x) - 2 \chi(R) \simeq 0 ,\end{equation}

\noindent where we have defined $R_{\rm GR} \equiv -T/M_{\rm
Pl}^{2} + R_{\rm vac}$. It should be understood that $R=R(x,R_{\rm
GR})$ in ($\ref{eq:i2}$), since the Ricci scalar will generically
be a function of both $x$ and $R_{\rm GR}$. Using a flat FRW
metric ansatz, we obtain

\begin{equation}\label{eq:3}  \chi'_{0} \left[x_{\lambda\lambda} +
\left[ 2(\log \chi_{0}')_{\lambda}  + 3 \bar{H} \right] x_{\lambda}\right]
+ (1+x) \Box \chi'_{0} +  {R(x,R_{\rm GR})) - R_{\rm GR} \over 3} +
{2\chi(R) - R(x)\chi'_{0} (1+x) \over 3} = 0 ,\end{equation}

\noindent where $\bar{H} = a_{\lambda}/a$. Now by using the
conditions $\chi'(R) \ll 1$ and $\chi(R)/R \ll 1$, Eq.
($\ref{eq:3}$) can be approximated as

\begin{equation}\label{eq:4}  \chi'_{0} \left[x_{\lambda\lambda} +
\left[ 2(\log \chi_{0}')_{\lambda}  + 3 \bar{H} \right] x_{\lambda}\right]
+  {R(x,R_{\rm GR}) - R_{\rm GR} \over 3} \simeq 0 .\end{equation}

\noindent The $\chi'_{0}$ function multiplying the
$x_{\lambda\lambda}$ and $x_{\lambda}$ terms is unimportant and
can be removed by a redefinition of the time coordinate. The final
step is to invert the expression $1+x = \chi'(R)/\chi'_{0}$ to
obtain $R = R(x,R_{\rm GR})$ and substitute this in
($\ref{eq:4}$). Then we can associate the last term on the
left-hand side of ($\ref{eq:4}$) with a potential gradient
$\partial V/\partial x = (R(x,R_{\rm GR})-R_{\rm GR})/3$.

The two models considered in this paper have particularly simple
$\chi(R)$ functions, and hence it is straightforward to invert the
expression $1+x = \chi'(R)/\chi'_{0}$ to obtain $R = R(x,R_{\rm
GR})$, giving an analytic form for $\partial V/\partial x$.
However, in general we will not be able to write $\partial
V/\partial x$ in terms of known functions. Another feature of the
two models under consideration is that all time dependence in
$\partial V/ \partial x$ drops out, making $dV/dx$ a function of
$x$ only (this requires a further redefinition of the time
coordinate in the HSS model). Generically, $\partial V/\partial x$
will depend on both $x$ and $R_{\rm GR}$.

To show that the singularity is a generic feature of these models,
we write $\partial V/ \partial x$ as

\begin{equation} {\partial V \over \partial R} = {1 \over 3}{\partial x \over
\partial R}(R - R_{\rm GR}) , \end{equation}

\noindent which, by using the definition of $x$, can subsequently be
written as

\begin{equation} {\partial V \over \partial R} = {1 \over 3}{\chi''(R)\over \chi'_{0}}
(R - R_{\rm GR}) .\end{equation}

\noindent By integrating this expression, we obtain the potential
as a function of $R$ and $R_{\rm GR}$,

\begin{equation}\label{eq:ui1} V(R,R_{\rm GR}) = {\chi'(R) \over 3\chi'_{0}}
(R - R_{\rm GR}) - {\chi(R) \over 3\chi'_{0}} + \lambda(R_{\rm GR}) , \end{equation}

\noindent where $\lambda(R_{\rm GR})$ is an arbitrary function of
$R_{\rm GR}$. From ($\ref{eq:ui1}$) and the expressions $\partial
V/ \partial x = (R - R_{\rm GR})/3$ and $1+x =
\chi'(R)/\chi'_{0}$, it is clear that for any model in which
$\chi'(R)R \to 0$ (more rigorously, $R\chi''(R)$ is integrable)
and $\chi(R) \to 0$ as $R \to \infty$, the potential will possess
the singular point $\partial V/\partial x \to \infty$ and $V \to
\lambda(R_{\rm GR})$ as $x \to -1$.

We note that the singularity occurs for models in which $F''(R) >
0$ at large curvatures, and hence is unrelated to the
Dolgov-Kawasaki instability \cite{d1}. The Dolgov-Kawasaki
instability corresponds to exponentially growing scalaron modes
that are generically present in models that satisfy $F''(R)<0$ in
some dynamically accessible regime. We only consider models for
which $F''(R) > 0$ for $R > R_{\rm vac}$.

\subsection{Determination of the Hubble parameter}

In obtaining $x$ and hence $R$ from ($\ref{eq:l10}$) and
($\ref{eq:hu127}$), it was assumed that the Hubble parameter could
be written as $H = H_{\rm GR} + \delta H$, where $\delta H \ll
H_{\rm GR}$ is small throughout the evolution and hence can be
neglected. To check that this assumption is valid, the $(0,0)$
component of the Einstein equations should also be solved for $H$
to check that it does not diverge when $R \to \infty$.

The $(0,0)$ component is given by

\begin{equation} \label{eq:n1} 18 H F''(R) \left(\ddot{H} +4 H \dot{H}\right) +
{F(R) \over 2} - 3 \left( \dot{H} + H^{2} \right) F'(R) = {\rho \over M_{\rm Pl}^{2}} ,
\end{equation}

\noindent which, for the AB and HSS models, can be expanded as

\begin{equation} \label{eq:n2}  18 H \chi''(R) \left(\ddot{H} +4 H \dot{H}\right) +
3H^{2} + {\chi(R) \over 2} - 3 \left( \dot{H} + H^{2} \right) \chi'(R) - {R_{\rm vac}
\over 4} = {\rho \over M_{\rm Pl}^{2}} , \end{equation}

\noindent for $R > R_{\rm vac}$. Eqn ($\ref{eq:n1}$) is the first
integral of the trace of the gravitational field equations, and is
a second order differential equation for the Hubble parameter.
However, due to the oscillatory behaviour of $R$, it is difficult
to solve ($\ref{eq:n2}$) over a significant dynamical range, since
the divergence in $R$ is extremely sensitive to the initial
conditions of $H$ and $\dot{H}$. This is a manifestation of the
singular behaviour of the Ricci scalar. However, we will be able
to solve ($\ref{eq:n1}$) for the regularized models presented in
section 3.

\subsection{Structure of a singularity with $F''(R)=0$ for a finite
$R$}

\noindent Thus, we have shown that the ``Big Boost" weak curvature
singularity (\ref{BB1},\,\ref{BB2}) arises whenever $RF''(R)$ is
integrable for $R\to\infty$ with $F''(R)>0$. Now, for
completeness, let us present the structure of an even weaker
singularity arising when $F''(R)$ becomes zero at some finite
value of $R$, $R=R_s$, so that the second stability condition
(\ref{stabil}) is marginally violated. For a generic case with
$F'''(R_s)\not= 0$, as follows from Eq. (\ref{trace-eq}), this
occurs at a finite moment of time $t=t_s$ when $a$, $H$ and $R$
remain finite, but $\dot R$ diverges $\propto |t-t_s|^{-1/2}$.
Thus, the scale factor has the following behaviour for $t\to t_s$:

\begin{equation}\label{ws}
a(t)=a_0+a_1(t-t_s)+a_2(t-t_s)^2 + a_3|t-t_s|^{5/2} + ...~.
\end{equation}
The metric is $C^2$, but not $C^3$, continuous across this
singularity, and there is no unambiguous relation between the
coefficients $a_3$ for $t<t_s$ and $t>t_s$.

Therefore, all that was said in section 2.1 regarding the ``Big
Boost" singularity also applies to this weak singularity. Namely,
its appearance in solutions of interest for cosmology results in
the loss of predictability of future Cauchy evolution. Therefore,
if we choose to remain inside the scope of $F(R)$ gravity, models
where the second of the conditions (\ref{stabil}) is even
marginally violated during evolution should be avoided.

\section{\label{sec:3}Avoiding the weak singularity and solving
the problems of $F(R)$ DE models}

In the previous section, we have reviewed the existence of the
weak ``Big Boost" singularity which occurs in general at a finite
redshift in the HSS and AB models, as well as in any other $F(R)$
model of present DE for which $F''(R)>0$ and $RF''(R)$ is
integrable for $R\to \infty$. Since the problem arises at large
curvatures, it is clear that $F(R) \approx R - const$ as $R \to
\infty$ is an inappropriate law for viable $F(R)$ DE
models.\footnote{This refers also to combined models of primordial
and present DE considered recently in \cite{ona,onb} which have
the same behaviour of $F(R)$ for $R\to\infty$. Generically these
models possess weak singularities either of the type
(\ref{BB1},\,\ref{BB2}), or of the type (\ref{ws}), since the
second of the stability conditions (\ref{stabil}) is violated for
them, too.} Under the less restrictive conditions $F''(R)>0$ and
$RF''(R)\to 0$ as $R\to\infty$, another difficulty, the first point
mentioned in the Introduction, arises; an unlimited growth of
the scalaron rest-mass $M_s=(3F''(R))^{-1/2}$ in the quasi-GR
regime (\ref{GR}). If $M_s$ exceeds $M_{\rm Pl}$, while $|R|$
remains less than $M_{\rm Pl}^2$, particles (scalarons) should
collapse to black holes from a naive physical point of view. More
formally, this means that loop quantum-gravitational corrections
to the tree action (\ref{S}) become dominant, so it may not be
used further in an effective quasi-classical theory. Thus, such
models of present DE are in general not compatible with the
standard early Universe cosmology including the correct BBN and
recombination.

Since the origin of these two difficulties is that
$F''(\infty)=0$, they can both be cured by a very simple change in
the HSS and AB models. Let us add an additional term, quadratic in
the Ricci scalar, to their $F(R)$ functions, so that $F''(\infty)$
becomes non-zero. In particular, we consider the following
functions,

\begin{eqnarray} \label{R-corr} \hat{F}_{\rm HSS} = F_{\rm HSS}(R) +
 {R^{2} \over 6M^{2}}, \qquad    \hat{F}_{\rm AB} = F_{\rm AB}(R) +
 {R^{2} \over 6M^{2}} , \end{eqnarray}

\noindent where $M$ is a mass scale coinciding with the scalaron
rest-mass whenever low curvature modifications to GR can be
neglected. There is no ``Big Boost" singularity in solutions of the
$R^2$-corrected HSS and AB models (\ref{R-corr}).

If we now return to the linearized analysis for these new models,
in which $R = R_{\rm GR} + \delta R$, then it follows from
Eq.\,(\ref{R-osc}) that the quadratic term has two important
effects on the dynamics of $\delta R_{\rm osc}$ -- the oscillating
component of $\delta R$. The first is that it introduces an upper bound
on the mass of the scalaron, $M_{\rm s}\leq M$, and hence limits
the frequency of oscillations of $R$ (as was noted in \cite{st}).
The second effect is to moderate the amplitude growth of $\delta
R_{\rm osc}$, which now goes $\propto a^{-3/2}$. Specifically,
going to the past, $\delta R_{\rm osc} /R_{\rm GR}$ decreases
throughout the matter and radiation eras beginning from the moment
when the $R^2$ correction in Eq.\,(\ref{R-corr}) becomes larger
than the non-GR term decaying with the growth of $R$. Hence
$\delta R_{\rm osc}$ remains a small perturbation throughout the
cosmological evolution, subject to it being small at present. Vice versa,
going to the future time direction, since $\delta R_{\rm osc}
/R_{\rm GR}$ grows until recently for the models (\ref{R-corr}),
it remains an open problem to explain its very small initial
amplitude in the early Universe. It is clear that the scalaron overabundance
problem noted in the Introduction has not yet been solved.

Let us now discuss possible values of the parameter $M$ in Eq.
(\ref{R-corr}). It should be sufficiently large in order to
satisfy the viability conditions presented in the Introduction. In
particular, it may not be $\sim \sqrt {R_{\rm vac}}\sim 10^{-33}$
eV as was considered in \cite{NO03,D04}, and even the values
discussed in \cite{DJJNST08,KM09} are not high enough to solve the
overabundance problem. A non-oscillating part of $\delta R$
induced by the $R^2$ correction to $F(R)$ becomes important when
$H(t)\sim M$. As a consequence of this, the lower limit
$M>10^{-2.5}$ eV which follows from the most recent laboratory
Cavendish-type experiment \cite{K07} is already sufficient for
this correction to GR to be negligible both during BBN and in the
center of neutron stars.\footnote{Still, as we shall see from
section 4, non-GR terms in Eq.\,(\ref{R-corr}) might become
important should the trace $T$ of the matter energy-momentum tensor
change sign and become positive inside neutron stars, as happens
in idealized $\rho = const$ solutions considered in
\cite{ma,KM09}. However, it is argued in \cite{la1} (see also
\cite{huu1}) that this does not occur inside realistic neutron
stars.}

However, taking $M$ close to this lower limit is incompatible with the
existence of any kind of inflation (not specifically driven by
$F(R)$ gravity) in the early Universe, which is required to solve
many other cosmological problems. Indeed, if $M \ll H_{\rm inf}$
where $H_{\rm inf}$ is the Hubble parameter during the last part
of inflation, quantum fluctuations of the scalaron, including
long-wave ones, are generated during inflation that generically
results in a large value of $\delta R_{\rm osc}$ after the end of
inflation, and can even lead to the existence of a second stage of
inflation driven by the scalaron itself \cite{KLS85}. Of course,
the scalaron is not stable and decays into pairs of particles and
antiparticles of all non-conformal quantum fields (this is a
particular case of the effect of particle creation in
gravitational fields). However, this process is sufficiently slow.
Even in the pure $R+R^2/6M^2$ model which does not describe the
present DE, the characteristic scalaron decay time is $\tau \sim
M_{\rm Pl}^2M^{-3}$ \cite{st1,S82,vk1,FTBM07}. Thus, one would
need $M\gg 10^5$ GeV to ensure the scalaron decay by the moment
when $H\sim 1$ s$^{-1}$ to avoid any problems with BBN. However,
we shall see in the next section that a non-trivial low-$R$
structure of $F(R)$ models describing present DE results in
slowdown of the scalaron decay after the end of inflation driven
by the scalaron itself. Since the aim of the present paper is to
find at least one $F(R)$ model of present DE satisfying all five
viability conditions formulated in the Introduction, we choose the
value of $N$ which is {\em sufficient} to solve the scalaron
overabundance problem. Namely, we take $M$ either larger than
$H_{\rm inf}$ in case inflation is produced by some other scalar
field, or $\approx 3.7\times 10^{13}\,(50/N)$ GeV if the scalaron
plays the role of an inflaton, as was noted in the Introduction.
In the former case, scalarons are practically not generated during
inflation, so $\delta R_{\rm osc}$ is zero after its end. In the
latter case, reheating after inflation appears to be very
non-trivial; it is studied in section 4.

Let us now return to the evolution of the $R^2$-corrected HSS and
AB models at recent redshifts. Although the above reasoning
suggests that the perturbative analysis is sufficient, it is also
interesting to investigate the effect of the quadratic term on the
scalaron potential. Performing the same steps as in section
\ref{sec:2}, the potentials are exhibited in fig.\ref{fig:3}. We
see that the singular point at which $V(x) \to {\rm const}$,
$dV/dx \to -\infty$ is no longer present, and we observe a regular
potential for all values of $x$ of physical interest.

\FIGURE{\hspace{10mm}\epsfig{file=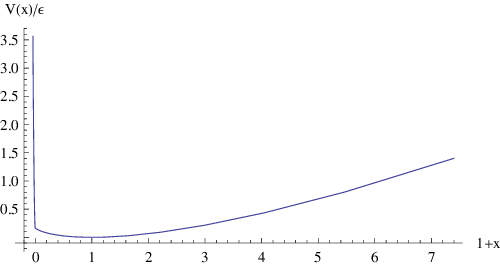,width=0.7\textwidth}\\
        \epsfig{file=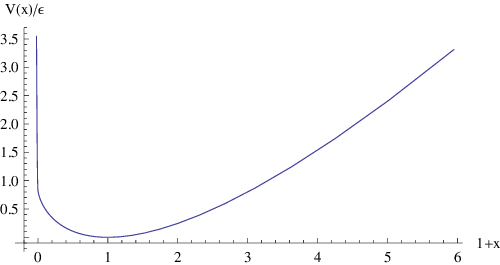,width=0.7\textwidth}
        \caption[]{\label{fig:3}The potential for the $R^2$-corrected AB
        model (top) and HSS model (bottom).
        We have taken $b = 4$, $R_{\rm GR} = 10 \times R_{\rm vac}$,
        $\delta \equiv \epsilon/M^{2} = 1.5\times 10^{-9}$  and
        $\alpha_{1} = 0$ for the AB model, and $n=2$, $c=1$,
        $R_{\rm GR} = 10\times R_{\rm vac}$, $\delta = 1.5 \times 10^{-9}$
        and $\alpha_{2} = 0$ for the HSS model. We see that $V(x) \to \infty$
        as $x \to -\infty$, and hence the singular behaviour has been removed
        from both models. Specifically there is no singularity at $x \to -1$,
        since both $V(x)$ and $dV/dx$ are now regular at this point. }

    }

\FIGURE{\hspace{15mm}\epsfig{file=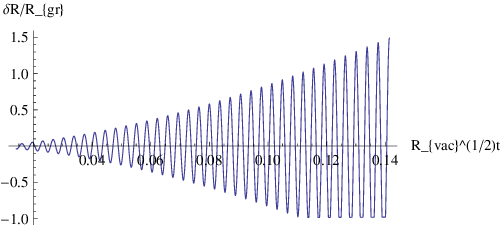,width=0.7\textwidth}\\
        \epsfig{file=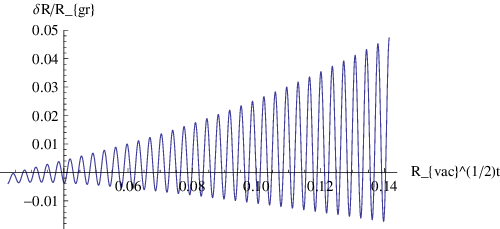,width=0.7\textwidth}
        \caption[]{\label{fig:7}The fractional difference $\delta R/R_{\rm GR}
        \equiv (R-R_{\rm GR})/R_{\rm GR}$ for the $R^2$-corrected AB model, obtained
        by numerically evolving the trace of the gravitational field equations
        using $\delta \equiv \epsilon/M^{2} = 4\times 10^{-8}$, and taking
        $R_{\rm GR} = 4/3t^{2}$ (that is, we evolve the Ricci scalar assuming matter
        domination). We take random initial conditions for $R$ and $dR/dt$, and use
        a dimensionless time coordinate $\hat{t} = R_{\rm vac}^{1/2}t$, where $t$ is
        the cosmological time. (a) and (b) differ only by the initial conditions placed on $R$ and $dR/dt$; in (a) we have perturbed $R$ significantly away from its General Relativistic value. We observe that the oscillatory component of the Ricci
        scalar decays to the past relative to $R_{\rm GR}$, as predicted in the text. }

    }

In fig.\ref{fig:7}, we have confirmed numerically that the
introduction of the $R^{2}$ term bounds the oscillations, which
now grow to the past at a slower rate than $R_{\rm GR}$ in the AB
model. To obtain these curves we have numerically evolved the
trace of the gravitational field equations, using the
$R^2$-corrected AB model and taking $\delta \equiv \epsilon/M^{2}
= 4\times 10^{-8}$. We have evolved equation ($\ref{trace-eq}$)
through the matter era, using a dimensionless time coordinate
$\hat{t} = R_{\rm vac}^{1/2}t$ and using random initial conditions
for $R$ and $dR/d\hat{t}$. In the AB and HSS models without the
$R^{2}$ term, it was found that $R$ could only be evolved over
very short timescales (evolving backwards over the matter era),
and only if we fine tuned initial conditions to approximately $R
\simeq R_{\rm GR}$. Now, we can choose random initial conditions
for $R_{\rm i}$, $\dot{R}_{\rm i}$ and $H_{\rm i}$, $\dot{H}_{\rm
i}$ and evolve backwards. In all cases we observe that the
oscillations of $R$ decay to the past, and no singularity is
present. Unfortunately, we cannot directly compare the results
obtained here with the corresponding functions $\delta R$ and
$\delta H$ in the uncorrected AB and HSS models. This is due to
the fact that the uncorrected models will generically evolve to a
singularity over much shorter timescales than presented here (see
ref.\cite{ap2} for a discussion of the difficulties associated
with numerically modelling the original AB and HSS models).

Summarizing, we have found a way to cure all three problems of the
HSS and AB models of present, low-curvature, DE which does not
destroy correct inflation, BBN and other advantages of the early
Universe cosmology. The approach consists of changing the behaviour of
$F(R)$ at $R\gg R_{\rm vac}$ according to Eq.\,(\ref{R-corr}) with
a very large value of the free parameter $M$ which should either
exceed the scale of inflation, or be equal to the concrete value
needed for scalaron driven inflation in $F(R)$ gravity. Still this
not the end of the story, one loophole remains which requires
further correction of the functions (\ref{R-corr}), now in the
range $R<R_{\rm vac}$ including negative values of $R$. This final
step in constructing a viable $F(R)$ DE model will be made in
section 4.

Let us finish this section with a comment on possible alternatives
to the large-$R$ behaviour (\ref{R-corr}). If instead a more
general term $M^{2-2m}R^m,~m>1,$ is added to $F_{HSS}$ or
$F_{AB}$, it can be checked that $M_s$, while growing with $R$,
never exceeds it, so it may not become larger than $M_{\rm Pl}$ if
$R$ is less than $M_{\rm Pl}^2$. Thus, there is no problem with
the unlimited growth of $M_s$. However, instead we face a new
difficulty: a possibility of the formation of a new space-like
curvature singularity with
\begin{equation}\label{BR}
a(t)\propto (t_s-t)^q~, ~~~q=\frac{(m-1)(2m-1)}{2-m}
\end{equation}
\noindent during evolution to the future which destroys all
subsequent evolution \cite{Sch89,MSS90,COM92}. For $m>2$, this
singularity is of the ``Big Rip" type, and the scale factor $a(t)$
becomes infinite at a finite moment of time.  For $1<m<2$, the
scale factor becomes zero at $t=t_s$. In both cases, it remains an
open problem if the singularity (\ref{BR}) can be avoided in
generic future Cauchy evolution after inflation. So, at present
the large-$R$ behaviour (\ref{R-corr}) of $F(R)$, up to
logarithmic in $R$ corrections, seems to be the only one free from
dangerous pathologies.

\section{\label{sec:4}Inflation and late time acceleration from
one $F(R)$ function}

As has been shown in the previous sections, to construct a viable
model of present DE in the scope of $F(R)$ gravity which satisfies
all the viability conditions and does not destroy previous
successes of the early Universe cosmology, one has not only to
choose the correct non-GR structure of $F(R)$ at low curvatures
$R\sim R_{\rm vac}$, like that in the HSS or AB models, but also to
modify high-$R$ behaviour of $F(R)$. Moreover, it appears
that the only high-$R$ behaviour which does not lead to problems
with new singularities is just that was originally proposed for
the scalaron driven $R^2$ inflation in the early Universe.
Therefore, it is natural to consider combined $F(R)$ models which
describe both primordial and present DE using one $F(R)$ function,
albeit one containing two greatly different characteristic mass scales.

It presents no problem to find a function $F(R)$ for which the
equation (\ref{dS}) defining de Sitter solutions has two or more
roots. The real issues are, first, to ensure that the inflationary
de Sitter solution is metastable, slow-roll and leads to the
correct spectrum of primordial perturbations and, second, that
there exist a sufficiently effective mechanism of reheating after
 inflation which transfers energy from scalarons into ordinary
matter and radiation and heats them to a high temperature long
before the beginning of a second de Sitter stage which we observe
now as the present acceleration of the Universe. This transition
between the two accelerating epochs in $F(R)$ gravity must be
carefully analyzed and the absence of singular points or
instabilities in the cosmological evolution has to be proved.

\subsection{\label{sec:n1}New problem}

When we begin to consider post-inflationary evolution in $F(R)$
gravity a new problem immediately arises, not only in the combined case when inflation is
scalaron-driven but also when inflation is produced by a
minimally coupled scalar field $\psi$, which requires further generalization of even the
corrected models (\ref{R-corr}). Namely, while $R$ is positive
during both inflation and the recent evolution of the Universe, it
becomes negative (and large) during each post-inflationary
oscillation of $\psi$. In particular, $T\equiv 3P -\rho =
\dot\psi^2>0$ at the moment when $V(\psi)=0$. It will be shown
below that the same occurs after scalaron-driven inflation. Thus,
the range of $R$ in the models (\ref{R-corr}), previously used for
$R>R_0$ only, has to be extended to negative values, and the
stability conditions (\ref{stabil}) should be satisfied for
negative $R$, too, at least up to values $R\sim - M^2$ which occur
during post-inflationary evolution.

However, this is not possible to achieve without a further change
of these models. In the case of the HSS model, both original and
the $R^2$-corrected one, first, one has to assume additionally
that $n$ is an integer to avoid non-analytical behaviour at $R=0$
(note that there is no such problem in the variant of this model
introduced in \cite{st}). A more serious problem is the appearance
of the weak singularity (\ref{ws}) at the points $R/R_{\rm vac}=
\pm \left((2n-1)/c(2n+1)\right)^{1/2n}$ where $F''(R)=0$. As shown
in section 2.4, it is not possible to predict deterministic Cauchy
evolution through it in the generic case (when $a_3\not= 0$).

In the case of the AB model, its function $F_{AB}(R)$ is analytic
and satisfies the conditions (\ref{stabil}) for all $R$. However,
$F'_{AB}(-\infty)=0$ and it approaches this limit exponentially
fast for low values of $|R|$: $F'(R) \sim \exp[2(R/\epsilon-b)]$
for $R<0,~|R| \gg \epsilon$. As a consequence of this, the
$R^2$-corrected AB model (\ref{R-corr}) acquires a point at which
$\hat F'(R_{0})=0$ for $R_{0} \sim -R_{\rm vac}$. As explained in
the Introduction, such points should be avoided because of the
formation of a generic anisotropic curvature singularity
\cite{H73,GS79}. Since this occurs at very low curvatures in this
case, this point is dynamically accessible during the
post-inflationary phase of the Universe evolution, where $R$
oscillates around the vacuum state. Hence, the $R^2$-corrected AB
model is not viable either.

Moreover, integrating the viability condition $F''(R)>0$ over the
interval $(-R,R)$ with $R\ll M^2$, we obtain that any viable
$F(R)$ DE model should have a non-zero $g$-factor
\begin{equation}\label{g-factor}
g=\frac{F'(R)-F'(-R)}{2F'(R)}~,~~~R_0\ll R\ll M^2~.
\end{equation}
\noindent Physically this means that the value of the effective
background Newton gravitational constant $G_{\rm eff}=G/F'(R)$ in the
quasi-GR regime (\ref{GR}) is larger for $R<0$ than for $R>0$. It
follows from the stability conditions (\ref{stabil}) that the
$g$-factor always lies in the range $0<g<1/2$.

\FIGURE{\epsfig{file=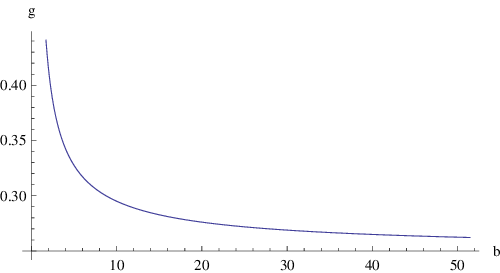,width=0.7\textwidth}
        \caption[]{\label{fig:11}The allowed parameter range of $g$
        and $b$. For any choice of $b$ and $g$ above the curve,
        the model ($\ref{eq:k1}$) has a stable de Sitter vacuum
        state. The curve is obtained by calculating the values of $g$
        and $b$ for which the functions $Q(R) \equiv RF'(R) - 2F(R)$
        and $dQ/dR$ are both zero.}
    }

\subsection{Resolution of the problem and the improved AB model}

This new problem can be solved and the problematic point $F'(R)=0$
for small $|R|$ can be avoided if we consider the improved,
$g$-extended $R^2$-corrected AB model, more concisely, the $gR^2$-AB model,

\begin{equation}\label{eq:k1} F(R) = (1-g) R + g \epsilon \log
\left[ {\cosh \left( R/\epsilon - b\right) \over \cosh b} \right] +
{R^{2} \over 6 M^{2}} ,\end{equation}

\noindent where we have introduced the new dimensionless parameter
-- $g$-factor, $0<g<1/2$. Note that $g=1/2$ corresponds to the
$R^2$-corrected AB model, and $g=0$ to the $R^{2}$ inflationary
model with $F(R) = R + R^{2}/6M^{2}$ \cite{st} which does not
present DE. The function ($\ref{eq:k1}$) corresponds to an
interpolation between two different gravitational constants, as
$F(R) \sim R$ for $R_{\rm vac} < R < M^{2}$ and $ F(R) \sim
(1-2g)R$ for $-M^{2} < R < \epsilon$, with a step at $R \sim
b\epsilon$.

\subsection{de Sitter attractors}

Like the AB and $R^2$-corrected AB models, this
new model can be expanded as $F(R) \approx R - R_{\rm vac}/2 +
\chi(R)$ for $R_{\rm vac} < R < M^{2}$, where now $R_{\rm vac} =
2g(b+\log [2\cosh b])\epsilon_{\rm AB}$ and $\chi(R) = g\epsilon
e^{2b} e^{-2R/\epsilon} + R^{2}/6M^{2}$. It has stable Minkowski
and de-Sitter vacuum states for an appropriate choice of $b$ and
$g$; in fig.\ref{fig:11} we have exhibited the allowed parameter
range. To obtain this curve we have used the fact that for
particular values of $b$ and $g$, the function $Q(R) \equiv RF'(R)
- 2F(R)$ will possess three zeros corresponding to vacuum states
of the model (Minkowski space, and two de Sitter vacua). The
values of $g$ and $b$ below the curve yield only one zero in
$Q(R)$ (Minkowski space), and those above will yield three. The
curve corresponds to the limiting case where $Q(R)$ possesses a
double zero, that is when $Q(R) = 0$ and $Q'(R) = 0$. Hence we
find that for any parameter choices of $b$ and $g$ above the
curve, there exists a stable de Sitter vacuum state. As we
increase $b$, the allowed range of $g$ increases, and as $b \to
\infty$, $g$ must satisfy $g \geq 0.25$. We also note that as $g
\to 0.5$, the condition $b \geq 1.6$ is required for a de Sitter
state. Finally, the model satisfies $F'(R) > 0$ for $R> -3M^{2}
(1-2g)$ and $F''(R) > 0$ for all $R$, and hence possesses no known
instabilities for $R > -3M^{2}(1-2g)$.

\subsection{Slow-roll inflation}

We now study the evolution of the model ($\ref{eq:k1}$), starting
from a high curvature slow roll epoch and evolving forwards in
time. We assume that the classical evolution of this model begins
with $H_{\rm i} \stackrel {<}{\sim}M_{\rm Pl}$ and $|\dot{H}|_{\rm
i} \sim M^{2} \ll H^{2}_i$. We analyze the $(0,0)$ component of
the Einstein field equations

\begin{equation} \label{eq:re1} 18 H F''(R) \ddot{H} + 72 F''(R) H^{2}
\dot{H} + {F(R) \over 2} - 3 \left( \dot{H} + H^{2} \right) F'(R) = 0 ,
\end{equation}

\noindent in order to study the behaviour of the Hubble parameter.
In the regime $ M^{2} < R < M_{\rm Pl}^{2}$, the function
($\ref{eq:k1}$) has the form $F(R) \approx R + R^{2}/6M^{2} -
R_{\rm vac}/2 + g \epsilon \exp[-2(R/\epsilon - b)]$, and the last
two terms are negligible (in particular, the low curvature
correction is exponentially suppressed). Therefore it is an
excellent approximation to use $F(R) = R + R^{2}/6M^{2}$ in
($\ref{eq:re1}$), which gives

\begin{equation} \label{eq:re2} \ddot{H} - {1 \over 2} {(\dot{H})^{2}
\over H} + 3H\dot{H} + {M^{2} \over 2}H = 0 . \end{equation}

\noindent The slow-roll evolution of ($\ref{eq:re2}$) (and
generalizations thereof) has been studied extensively \cite
{st3,vk1,mj1,mj2}, and it has been shown that for $H^{2} \gg
|\dot{H}|$, $H|\dot{H}| \gg |\ddot{H}|$, the Hubble parameter has
the following form

\begin{equation} H(t) \approx H_{\rm i} - {M^{2} t \over 6} ,\end{equation}

\noindent where we have taken $t=0$ at the beginning of the
evolution and $H_{\rm i} \stackrel {<}{\sim}M_{\rm Pl}$. Slow roll
ends when $H \sim M$. Following this, we have an epoch in which
the both the Hubble parameter and Ricci scalar oscillate with high
frequency around a stable ground state. In the following section,
we consider the oscillations of the Ricci scalar in detail, and
how they may reheat the Universe.

Let us show now that all other scalaron-driven inflationary models
in $F(R)$ gravity which produce primordial spectra with other
values of $n_s$ and $r$ are, in a sense, close to the $R^2$ model.
Indeed, there may be two kinds of inflationary models. In the
first one which is the analogue of chaotic inflation \cite{L83} in
GR, inflation occurs over a wide range of $R$. Then it follows
from Eq.\,(\ref{eq:re1}) that we need $F(R)\approx R^2A(R)$ for
$R\to\infty$ with $A(R)$ being a slowly varying function of $R$,
namely
\begin{equation}\label{chaotic}
|A'(R)|\ll \frac{A(R)}{R}~,~~~|A''(R)|\ll \frac{A(R)}{R^2}~,
\end{equation}
\noindent for this kind of inflation to take place. Thus, these
models are indeed close to the $R^2$ one.

In the second case, inflation occurs around some fixed root
$R=R_1$ of Eq.\,(\ref{dS}) -- an analogue of the ``new"
inflationary model \cite{L82,AS82} in GR. Then, taking into
account that the inequality (\ref{dSstab}) should be satisfied
only marginally for metastability of the de Sitter solution, we
get
\begin{equation}\label{new}
F'(R_1)= \frac{2F(R_1)}{R_1}~,~~~ F''(R_1)\approx
\frac{2F(R_1)}{R_1^2}~.
\end{equation}
Thus, these models are close to the $R^2$ one near the point
$R=R_1$.

 Finally in this section, we exhibit the potential of the scalaron
in the Einstein frame in fig.\ref{fig:e13}. We observe the
metastable de-Sitter point, and the stable Minkowski vacuum.

\FIGURE{\epsfig{file=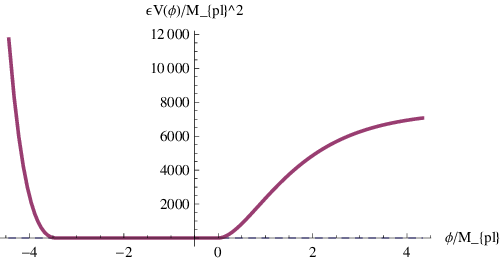,width=0.8\textwidth}\\
        \epsfig{file=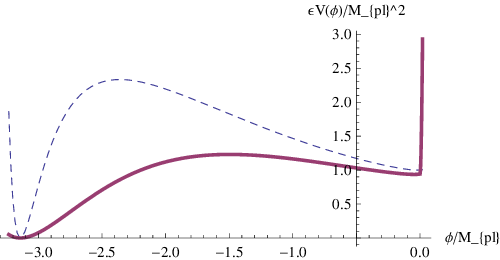,width=0.8\textwidth}
        \caption[]{\label{fig:e13}The Einstein frame potential for the
        $gR^2$-AB model (solid) and the original AB model (dashed).
        To obtain the potential we have used $g=0.47$, $\delta =
        \epsilon/M^{2} = 10^{-4}$ and $b=2$. We observe two vacuum
        states; one late time de Sitter attractor and a flat Minkowski
        vacuum state. Slow-roll occurs for $\phi \gg M_{\rm Pl}$. We
        note that in the original model, the scalar field satisfies
        $\phi <0$ and the singularity corresponds to the point $\phi
        \to 0$. This singularity is removed in the $gR^2$-AB model
        by mapping the singular point $R \to \infty$ in the Jordan
        frame to $\phi \to \infty$ in the Einstein frame.}

    }

\subsection{Reheating}

We now study the reheating epoch for this class of models (a
discussion of reheating mechanisms for modified gravity models can
be found in, for example, \cite{S82,vk1,FTBM07}). For the $R^{2}$
inflationary model, immediately following slow roll we have $H^{2}
\sim |\dot{H}| \sim M^{2}$, and the Ricci scalar undergoes damped
harmonic oscillations around $R=0$ with frequency $\omega = M$ and
$|R| \ll M^{2}$. In this section, we consider the evolution of $H$
for the model in question, taking initially $H^{2} \sim |\dot{H}|
\sim M^{2}$. We find behaviour that is dramatically different to
that of the $R^{2}$ model, suggesting that the modifications to
General Relativity at $R \sim 0$, which were introduced to induce
late time acceleration, also have a significant effect on the
dynamics of the Ricci scalar in this early epoch of the Universe.
In any unified model of inflation and present DE there must be an
efficient reheating mechanism, otherwise the Universe will relax
to its late time de Sitter attractor without first undergoing
epochs of radiation and matter domination.

To study reheating for the model ($\ref{eq:k1}$) we again consider
the (0,0) field equation, which can be written in terms of
dimensionless parameters $\hat{H} = H/M$, $\hat{R} = R/M^{2}$,
$\delta \equiv \epsilon/M^{2} \ll 1$ and dimensionless time
$\hat{t} = Mt$,

\begin{eqnarray}\label{eq:rt1} \hat{H} \hat{H}'' - {(\hat{H}')^{2}
\over 2} + 3 \hat{H}^{2} \hat{H}' + {(1-g) \over 2} \hat{H}^{2} -
{g \over 2} (\hat{H}'+\hat{H}^{2}) \tanh \left[{ \hat{R} \over
\delta} - b \right] \\ \nonumber + {g \delta \over 12} \log \left[
{\cosh( \hat{R} /\delta - b) \over \cosh (b)} \right]  + {3 g\over
\delta} {\rm sech}^{2} \left[{\hat{R} \over \delta} - b \right]
(\hat{H} \hat{H}'' + 4\hat{H}^{2} \hat{H}') = 0 ,\end{eqnarray}

\noindent where primes now denote derivatives with respect to
$\hat{t}$. Initially, we will solve equation ($\ref{eq:rt1}$),
neglecting any effects due to gravitational particle production
(that is, we will neglect the backreaction of created particles on
the dynamics of $\hat{H}(\hat{t})$). The issue of backreaction
will be tackled in section \ref{sec:r1}.

\subsubsection{Evolution of $H(t)$ without backreaction}

To begin, we solve ($\ref{eq:rt1}$) numerically. We are
considering the epoch immediately following slow roll, so we take
as initial conditions $\hat{H}'_{\rm i} = -1$ and $\hat{H}_{\rm i}
= 1$ and evolve in $\hat{t}$ (we will begin our evolution at
$\hat{t} = 0$). The results are exhibited in fig.\ref{fig:rt1} for
the Hubble parameter and fig.\ref{fig:rt2} for the absolute value
of the Ricci scalar, where we have taken $\delta = 1.5 \times
10^{-8}$, $b=3$ and $g=0.45$. We note that for realistic choices
of $M$ and $\epsilon$, $\delta$ will be many orders of magnitude
smaller, however numerically we are restricted to $\delta \sim
10^{-8}$. As can be seen in fig.\ref{fig:rt1}, $\hat{H}$ evolves
through a number of distinct regimes, which we discuss below.

Initially, $\hat{R} \sim 1$, so we can use $F(R) \approx R +
R^{2}/6M^{2}$, and the Hubble parameter behaves as in the $R^{2}$
inflationary model (that is, $\hat{H}$ rapidly decreases to
$\hat{H} \ll 1$ over timescales $\hat{t} \sim {\cal O}(1)$). This
region corresponds to $\hat{t} \lesssim 1$ in
figs.\ref{fig:rt1},\ref{fig:rt2}. In this regime $\hat{R} \gg
\delta$, and hence the value of $\delta$ will have no significant
effect on the dynamics of $\hat{H}$ and $\hat{R}$.

After this (very short) initial period, we observe a significant
change in the gradient of $\hat{H}$. In this regime, we have
$\hat{H}^{2} \gg \delta$, $|\hat{H}'| \gg \delta$ but $\hat{R} =
6\hat{H}' + 12 \hat{H}^{2} \gtrsim \delta$ (that is, $\hat{H}'
\approx -2\hat{H}^{2}$.) The exact value of $\hat{R}$ will depend
on the specific function $F(R)$ being considered, and corresponds
to its value when the large curvature contribution to the scalaron
mass is approximately equal to the low curvature terms; for the
model constructed here when $1/3M^{2} \sim (g/\epsilon) {\rm
sech}^{2}[R/\epsilon - b]$.

To analytically model the behaviour of $\hat{H}$ in this regime,
we look for a solution to ($\ref{eq:rt1}$) of the form
$\hat{H}(\hat{t}) = \hat{H}_{0}(\hat{t}) + \delta
\hat{H}_{1}(\hat{t}) + {\cal O}(\delta^{2})$. In doing so we
obtain

\begin{eqnarray}\nonumber &\delta& \left( \hat{H}_{0} \hat{H}_{0}''
- {(\hat{H}_{0}')^{2} \over 2} + 3 \hat{H}_{0}^{2} \hat{H}_{0}' +
{(1-g) \over 2} \hat{H}_{0}^{2} - {g \over 2}
(\hat{H}_{0}'+\hat{H}_{0}^{2})
\tanh \left[ 6\hat{H}_{1}'+24\hat{H}_{0}\hat{H}_{1} - b \right]\right)  \\
\nonumber &+& 3 g {\rm sech}^{2}\left[6\hat{H}_{1}'+24\hat{H}_{0}
\hat{H}_{1} - b \right](\hat{H}_{0} \hat{H}_{0}'' + 4\hat{H}_{0}^{2}\hat{H}_{0}') \\
 \label{eq:rt51}  &+& 3 g \delta {\rm sech}^{2}\left[6\hat{H}_{1}'+
 24\hat{H}_{0}\hat{H}_{1} - b \right] (\hat{H}_{0} \hat{H}_{1}''+\hat{H}_{1}
 \hat{H}_{0}'' + 8\hat{H}_{0}\hat{H}_{1}\hat{H}_{0}' + 4\hat{H}_{0}^{2}
 \hat{H}_{1}') = 0 ,\end{eqnarray}

\noindent where we have anticipated that $\hat{R}_{0} =
6\hat{H}'_{0} + 12 \hat{H}^{2}_{0} = 0$ at zeroth order in the
$\tanh[\hat{R}/\delta-b]$ and ${\rm sech}^{2}[\hat{R}/\delta-b]$
terms (we will confirm that this assumption is valid below). At
order ${\cal O}(1)$ and ${\cal O}(\delta)$ we arrive at

\begin{eqnarray} &{\cal O}(1)& \qquad \hat{H}_{0} \hat{H}_{0}'' +
4\hat{H}_{0}^{2}\hat{H}_{0}' = 0, \\ &{\cal O}(\delta)& \qquad
\hat{H}_{0} \hat{H}_{1}''+\hat{H}_{1} \hat{H}_{0}'' + 8\hat{H}_{0}
\hat{H}_{1}\hat{H}_{0}' + 4\hat{H}_{0}^{2}\hat{H}_{1}' =
{\cosh^{2} \left[6\hat{H}_{1}'+24\hat{H}_{0}\hat{H}_{1} - b
\right] \over 3g} \times \\ \nonumber  & &
\left({(\hat{H}_{0}')^{2} \over 2} - \hat{H}_{0} \hat{H}_{0}'' - 3
\hat{H}_{0}^{2} \hat{H}_{0}' - {(1-g) \over 2} \hat{H}_{0}^{2} +
{g \over 2} (\hat{H}_{0}'+ \hat{H}_{0}^{2}) \tanh \left[
6\hat{H}_{1}'+24\hat{H}_{0}\hat{H}_{1} - b \right] \right)
,\end{eqnarray}

\noindent and hence at ${\cal O}(1)$, $\hat{H}_{0}(\hat{t}) = 1 /
(2\hat{t} + \alpha_{0})$, where $\alpha_{0}$ is an integration
constant. At zeroth order we have $a(\hat{t}) \propto
\hat{t}^{1/2}$, $\hat{H}(\hat{t}) \sim 1/2\hat{t}$ and $\hat{R}
=0$. This is in agreement with our numerical results; in
fig.\ref{fig:rt7} we have exhibited $\hat{H}$ as calculated
numerically and $\hat{H} = 1/2t$ logarithmically as a function of
$\log[t]$, over the range $\log[t] = (\log[1],\log[180])$, and we
see a close agreement between our analytic and numerical results.
The number of scale factor e-folds during this period is given by
$N_{1} \sim -\ln (1-2g)$.

\FIGURE[t]{\epsfig{file=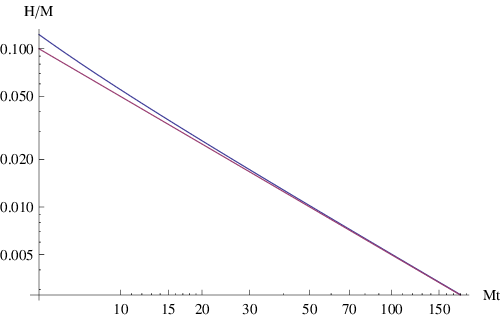,width=0.7\textwidth}
        \caption[]{\label{fig:rt7}exhibits $\hat{H}$ (solid),
obtained by numerically solving ($\ref{eq:rt1}$), and $1/2\hat{t}$
(dashed) plotted logarithmically against $\hat{t}$. We see that
over the range $\hat{t} = (5,180)$, $\hat{H} \sim 1/2\hat{t}$, in
agreement with our analytic result.}
    }

\FIGURE{\epsfig{file=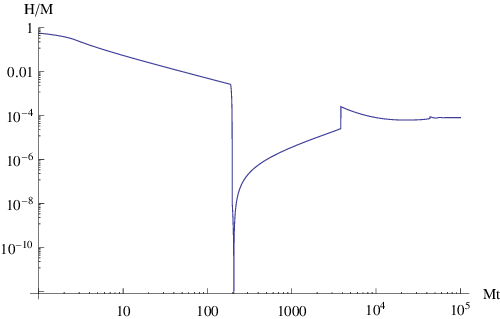,width=0.7\textwidth}
        \caption[]{\label{fig:rt1}The behaviour of $\hat{H}$
using the parameters $\delta = 1.5 \times 10^{-8}$, $b=3$,
$g=0.45$, and neglecting backreaction. We observe a number of
distinct regimes; for $0<\hat{t}<2$ the Hubble parameter is
exiting the slow roll regime. Following this is an epoch for which
$\hat{H} \sim 1/2\hat{t}$. The abrupt change in $\hat{H}$ at
$\hat{t} \sim 180$ corresponds to a spike in the evolution of
$\hat{R}$. Following this $\hat{H} \sim \delta^{1/2}$ and $a(t) \sim
{\rm const}$. Once $\hat{H}$ spikes once more at $\hat{t} \sim
5000$, the Hubble parameter has completed one full oscillation.}
    }

\FIGURE{\epsfig{file=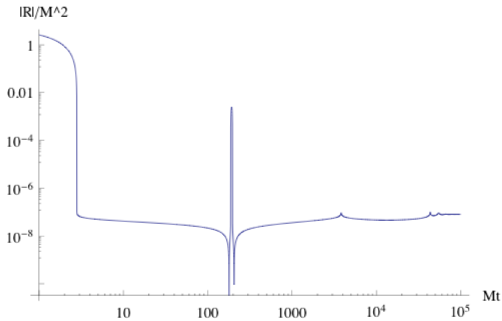,width=0.7\textwidth}
        \caption[]{\label{fig:rt2}The behaviour of the Ricci
scalar $\hat{R}$ using the parameters $\delta = 1.5 \times
10^{-8}$, $b=3$, $g=0.45$, and neglecting backreaction. We observe
the Ricci scalar evolving from a slow-roll de Sitter state to
$\hat{R} \gtrsim \delta$, where the low curvature corrections to
GR dominate. $\hat{R}$ is typically $\sim \delta$ throughout this
epoch, although it periodically exhibits spikes.}
    }

Reheating in this class of models will begin at this stage in the
evolution of $\hat{H}$, via gravitational particle production due
to the abrupt change in the Ricci scalar (in our numerical
calculation this occurs at $\hat{t} \sim 2$, as shown in
fig.\ref{fig:rt2}). Prior to this, the Ricci scalar is
approximately in a de Sitter phase, which then abruptly evolves to
a radiation-like Universe with $R \sim 0$. The main contribution
to the number of particles and antiparticles of all non-conformal
matter quantum fields created after the end of inflation occurs
during this (almost) discontinuous change in $R$. Using either the
expression for the rate of creation of the density of scalar
particles and antiparticles $n$ in the massless limit in a FRW
background \cite{zls,vk1}
\begin{equation}\label{rate}
\frac{d(a^3n)}{a^3\, dt}=\frac{(1-6\xi)^2}{576\pi}\, R^2~,
\end{equation}
which for $\xi = 0$ is valid for gravitons (although the rate should be doubled in this case due to two polarization states) and the longitudinal
component of vector bosons, too,\footnote{Strictly speaking,
Eq.\,(\ref{rate}) is derived in the limit $R^2\ll
R_{\mu\nu}R^{\mu\nu}$.} or results of the papers \cite{stz,fd}, we
arrive to the following estimate of the total number and energy
density of created ultrarelativistic particles:
\begin{eqnarray}\label{creation}
n \sim xH_{\rm r}^{3} \left({a_{\rm r}\over a}\right)^{3}~,~~~
\rho_{\rm rad} = x H_{\rm r}^{4}\left({a_{\rm r} \over
a}\right)^{4} ,\end{eqnarray}

\noindent where $H_{\rm r}$ is the value of the Hubble parameter
at which the discontinuous change in $R$ occurs (for our model
$H_{\rm r} \sim M$), $x$ is a dimensionless parameter of order $x
\sim 10^{-2}$ and $a_{\rm r}$ is the value of the scale factor at
this point. The energy density of created particles will be
initially subdominant, but may grow to have a significant
backreaction effect on the dynamics. For the remainder of this
section we neglect the effect of $\rho_{\rm rad}$ on the evolution
of $\hat{H}$ and consider backreaction in the following section.

Over the regime $\hat{H} = 1/2\hat{t} + {\cal O}(\delta)$
discussed above, the Ricci scalar is given by $\hat{R} \simeq 6
\delta \left( \hat{H}'_{1}+ 4\hat{H}_{0}H_{1} \right)$, which
decreases from $\hat{R} \gtrsim \delta$ to $\hat{R} \sim 0$, at
which time the $(1/\delta)\, {\rm sech}^{2}(\hat{R}/\delta - b)$
term in ($\ref{eq:rt1}$) no longer dominates. At this point, we
can use ${\rm sech}^{2}(\hat{R}/\delta - b) \simeq 0$ and
$\tanh(\hat{R}/\delta - b) \simeq -1$, in which case
($\ref{eq:rt1}$) becomes

\begin{equation} \hat{H}\hat{H}'' - {(\hat{H}')^{2} \over 2} +
3\hat{H}^{2} \hat{H}' + {(1-2g)\over 2}\hat{H}^{2} \simeq 0 ,
\end{equation}

\noindent which has solution $\hat{H} \simeq A_{0} \sin^{2}
(\sqrt{1-2g}\hat{t}/2)$ (that is, $\hat{H}$ oscillates on
timescales of order $\hat{t} \sim 1$, as in the $R^{2}$
inflationary model). This corresponds to the spike in $\hat{R}$ at
$\hat{t} \sim 200$, with amplitude $\hat{R} \gtrsim -1$. However,
unlike in the $R^{2}$ inflationary model, $\hat{R}$ only completes
one half of an oscillation before we once again enter a regime
where the $(1/\delta)\, {\rm sech}^{2}(\hat{R}/\delta - b)$ term
dominates.

Following the spike in $\hat{R}$, $\hat{H}$ does not decay like
$\hat{H} \propto \hat{t}^{-1}$. Instead, $\hat{H}^{2}, \hat{H}'
\lesssim \delta$, and the scale factor $a(t)$ is approximately
constant over this regime. This is confirmed numerically in
fig.\ref{fig:rt3}, where we observe a plateau in the evolution of
the scale factor. During this period, the number e-folds will be
$N_{2} \approx 0$, and $\hat{H}$ and $\hat{R}$ will grow until
$\hat{R} > \delta$.

Following this, $\hat{R}$ will once again produce a spike (with
amplitude $\hat{R} \lesssim 1$), completing one full oscillation
of $\hat{H}$. Numerically we can only observe one complete
oscillation before $\hat{R}$ reaches its final state $\hat{R}
\gtrsim \delta$, however this is simply a consequence of being
unable to choose a sufficiently small value of $\delta$. Finally,
we note that if we average $\hat{H}$ over an even number of
oscillations, we find $\langle \hat{H} \rangle = 1/3\hat{t}$, as
one might expect for a period of kination (domination of a
massless scalar field).

We can summarize the dynamics of $\hat{H}$ as follows; the Hubble
parameter periodically undergoes (almost) discontinuous jumps
between periods when $\hat{H} = 1/2\hat{t}$ and $\hat{H} \sim
\delta^{1/2}$. The duration of these periods in terms of $\ln t$ is
determined by the non-zero $g$-factor (\ref{g-factor}) and is
equal to $\approx -2\ln (1-2g)$ and $-\ln (1-2g)$ respectively.
It is clear that such behaviour of $\hat{H}$ and $\hat{R}$ is
markedly different to the standard reheating dynamics, and we
expect that the low curvature modifications to General Relativity
will leave unique observational imprints. In particular, our
reheating mechanism is less efficient than the pure $F(R) = R +
R^{2}/6M^{2}$ model, and we expect a significantly lower reheat
temperature. Additionally, the average expansion rate after
inflation is slower than $a(t) \propto t^{1/2}$, and therefore the
number of e-foldings, $N$, should be $\sim 70$, larger than in
standard inflationary models and the same that would occur if all
inflation proceeded at $H=M_{Pl}$ . Indeed, by comparing the
energy density of created particles (\ref{creation}) to $H^2$, we
see that they become equal at
\begin{equation}\label{reh}
t=t_{\rm reh}\sim x^{-3/2}M^{-4}M_{\rm Pl}^3 \sim 10^{-18}~{\rm s}
\end{equation}
after the end of inflation (assuming the value $M\approx 3\times
10^{13}$ GeV needed to fit the amplitude of observed curvature
fluctuations with $N=70$). If $g=0.45$, then about 7 complete
non-linear oscillations of $R$ have occurred by this moment. If
the radiation component has been already thermalized at $t=t_{\rm
reh}$ due to interactions between particles, then its
temperature is $T(t_{\rm reh})\sim 10^6$ GeV which is sufficiently
large (however, in principle, it may thermalize significantly earlier
while being a sub-dominant component). At $t=t_{\rm reh}$, the comoving scale which was equal to
the Hubble radius at the end of inflation is given by
$M^{-1}(Mt_{\rm reh})^{1/3}=x^{-1/2}M^{-2}M_{\rm Pl}=
x^{1/4}M_{\rm Pl}^{-1/2}t_{\rm reh}^{1/2}$. Therefore, up
to a factor of a few, it coincides with the comoving
scale which is equal to the Planck length at the Planck time
in a universe which is radiation-dominated at subsequent times.
As a result, irrespective of the fact that if the
thermodynamic equilibrium in the radiation component is reached
before or after $t_{\rm reh}$, this scale coincides with the
characteristic thermal length of present CMB photons with
temperature $T_{\gamma}=2.725$ K, up to a purely numerical factor
depending mainly on an effective number of species at the moment
when the equilibrum has been achieved. This explains why $N\approx 70$
for our model. In the next subsection, we
consider backreaction of created particles of a FRW background
numerically and in more detail.

Hence, due to this change in $N$, the index (slope) of the power
spectrum of primordial scalar (density) perturbations $n_{\rm s}$
is slightly higher, $n_{\rm s} =1-2/N \simeq 0.97$, in our model which
combines both inflation and present DE using one $F(R)$ function
(\ref{eq:k1}), as compared to the $R+R^2/6M^2$ model describing
inflation only. However, this distinctive and observable
prediction is degenerate, since it may be changed by introducing
loop corrections to the large curvature $R^{2}$ term of another
purely inflationary model, making it of the type (\ref{chaotic}).

\FIGURE{\epsfig{file=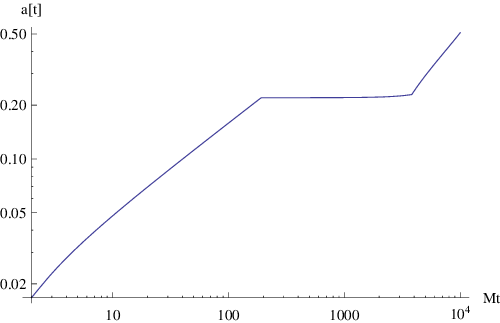,width=0.7\textwidth}
        \caption[]{\label{fig:rt3}The behaviour of $a(t)$, normalized
        such that $a(\hat{t}_{i})=2 \times 10^{-2}$. We observe a
        plateau in the evolution of the scale factor, as predicted in
        the text.}
    }

\subsubsection{\label{sec:r1}Effect of backreaction}

As pointed out in the previous section, at the beginning of the
reheating epoch $\hat{R}$ undergoes an almost discontinuous change
from a de Sitter phase to $\hat{R} \gtrsim \delta$, and particle
production occurs mainly during this period. To take into account
the backreaction of these particles on the evolution of the Hubble
parameter, we must now solve the equation

\begin{eqnarray} \label{eq:rt100} && \hat{H} \hat{H}'' - {(\hat{H}')^{2}
\over 2} + 3 \hat{H}^{2} \hat{H}' + {(1-g) \over 2} \hat{H}^{2} -
{g \over 2}
(\hat{H}'+\hat{H}^{2}) \tanh \left[{ \hat{R} \over \delta} - b \right]  \\
 \nonumber && + {g \delta \over 12} \log \left[ {\cosh( \hat{R}/\delta - b)
 \over \cosh (b)} \right] + {3 g \over \delta} {\rm sech}^{2}\left[{\hat{R}
 \over \delta} - b \right] (\hat{H} \hat{H}'' + 4\hat{H}^{2}\hat{H}') =
 {\rho_{\rm rad} \over 6M^{2} M_{\rm Pl}^{2}} \simeq {x M^{2}a_{\rm r}^{4}
 \over 6M_{\rm Pl}^{2} a^{4}} .\end{eqnarray}

\noindent For future convenience we define the dimensionless
parameter $\hat{\rho}_{\rm rad} \equiv \rho_{\rm rad}/M^{4}$.
Initially, the energy density $\rho_{\rm rad}$ of produced
particles will be subdominant to the background evolution of
$\hat{H}(\hat{t})$, driven by the scalaron. Numerically, we will
take the radiation component to be $\rho_{\rm i}/M^{2}M_{\rm
Pl}^{2} \sim 10^{-2} \hat{H}^{2}_{\rm i}$, and $xM^{2}/6M_{\rm
Pl}^{2} = 2 \times 10^{-4}$ (one must be careful to preserve the
hierarchy between $\hat{H}$, $\hat{\rho}_{\rm rad}$ and $\delta$,
as when $\hat{\rho}_{\rm rad}$ is of order $\delta$, our solution
will cease to be physically relevant.)

\FIGURE{\epsfig{file=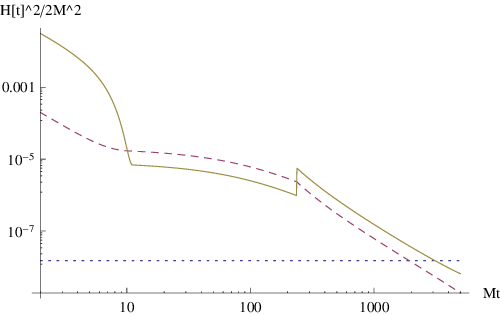,width=0.7\textwidth}
        \caption[]{\label{fig:rt5}The behaviour of $\hat{H}^{2}/2$
        (solid) and $M^{2}\hat{\rho}/6M_{\rm Pl}^{2}$ (dashed). We
        see that $\hat{H}^{2}/2$ oscillates around $\hat{\rho}$,
        and satisfies $\hat{H}^{2} \gg \delta$ throughout the
        reheating epoch ($\delta$ is also exhibited (dotted)).  }
    }

\FIGURE{\epsfig{file=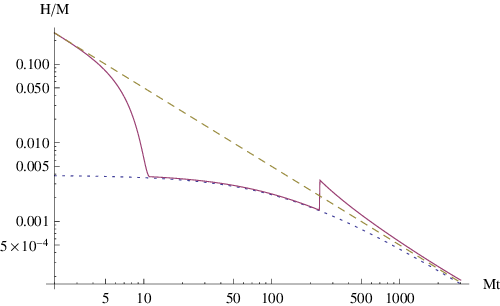,width=0.7\textwidth}
        \caption[]{\label{fig:rt6}The Hubble parameter during one
        complete oscillation (solid). We have also exhibited the two
        approximate analytic solutions as calculated in the text;
        $\hat{H} = 1/2\hat{t}$ (dashed) and $\hat{H} = 1/(2\hat{t}
        + \alpha)$ (dotted).}
    }

We begin our numerical evolution at the point where the radiation
energy density is produced, so $M^{2}\hat{\rho}_{\rm i}/6M_{\rm
Pl}^{2} = x M^{2} / 6M_{\rm Pl}^{2} = 2 \times 10^{-4} $, and use
initial conditions $2\hat{H}_{\rm i}^{2} = -\hat{H}'_{\rm i}$.
Taking $\delta = 1.5 \times 10^{-8}$, $b=3$ and $g=0.45$, we
present $\hat{H}^{2}/2$ and $M^{2}\hat{\rho}/6M_{\rm Pl}^{2}$ in
fig.\ref{fig:rt5}. We note that the observed behaviour of
$\hat{H}$ is significantly different to that of the previous
section, however our conclusions will remain essentially the same.
As before, we observe one complete oscillation of $\hat{H}$ before
$\hat{\rho} \sim \delta$. We now discuss the various regimes over
the course of this oscillation.

Initially, $\hat{H}$ behaves in a very similar manner to the
previous section; the radiation is subdominant and $\hat{H} =
1/2\hat{t}$, $\hat{\rho} \propto \hat{t}^{-2}$. During this time
$\hat{R}$ evolves from $\hat{R} \gtrsim \delta$ to $\hat{R} \sim
0$. Then, at the point $\hat{R} \sim 0$, the Ricci scalar
undergoes half an oscillation, spiking at $\hat{R} \gtrsim -1$
(again, as before). However, $\hat{H}$ dies not oscillate to
$\hat{H} = 0$ as in the previous section, but rather $\hat{H}^{2}$
oscillates around $M^{2}\hat{\rho}/3M_{\rm Pl}^{2}$. We see that
by incorporating $\hat{\rho}$ into the dynamics, $\hat{H}$ now
satisfies $\hat{H} \gg \sqrt{\delta}$ throughout the reheating
epoch.

All that remains is to consider the behaviour of $\hat{H}$ and
$a(t)$ over the second half of the oscillation of $\hat{H}$.
Following the (almost) discontinuous change in $\hat{H}$ at
$\hat{t} \sim 10$, the Hubble parameter satisfies $\hat{H} \gg
\sqrt{\delta}$ and $\hat{H}' \gg \delta$, however $\hat{R} \gtrsim
\delta$. By using the same perturbative analysis as before, we
must conclude that $\hat{H}$ is given by $\hat{H} = 1/(2\hat{t} +
\alpha)$, where $\alpha$ is an integration constant. The value of
this constant can be obtained numerically, and depends on the
ratio of $H^{2}_{\rm i}$ and $M^{2}\hat{\rho}/M_{\rm Pl}^{2}$. For
realistic values of $M$ and $M_{\rm Pl}$ we expect that $\alpha$
will be much greater than $\hat{t}$, and hence $\hat{H} \sim
\alpha^{-1}$ and $a(t) \approx {\rm const}$ during this period.
The number of e-foldings over this regime will be $N_{2} \approx
0$, as before. During this time $\hat{R}$ grows from $\hat{R} \sim
0$ to $\hat{R} \gtrsim \delta$, at which point we observe another
spike in $\hat{R}$, thus completing one full oscillation of
$\hat{H}$. In fig.\ref{fig:rt6} we have exhibited the Hubble
parameter and the two analytic approximations $1/2\hat{t}$ and
$1/(2\hat{t}+\alpha)$ derived above; we see that our results are
in agreement with the numerical solution.

Although the evolution of $\hat{H}$ is significantly modified when
we incorporate backreaction, our conclusions remain essentially
unchanged. $\hat{H}$ now undergoes periodic oscillations around
$3\hat{H}^{2} \sim M^{2}\hat{\rho}/M_{\rm Pl}^{2}$; the first half
of this oscillation is characterized by the behaviour $\hat{H} =
1/2\hat{t}$, whereas its behaviour over the second half is given
by $\hat{H} = 1/(2\hat{t}+\alpha) \approx 1/\alpha \ll 1$. As in
the previous section, we find that the reheating mechanism is less
efficient than the standard $R + R^{2}/6M^{2}$ model, and the
averaged Hubble parameter will evolve at a slower rate than
$1/2\hat{t}$, so our conclusions regarding the total number of
e-foldings and the consequent increase in the value of $n_{\rm s}$
will persist.

\subsection{Cosmological evolution}

Following the reheating epoch of the Universe, we expect that the
model ($\ref{eq:k1}$) reproduces the standard cosmology, that is
it evolves from a radiation dominated epoch to matter domination,
with the final state of the Universe being the de Sitter vacuum.
To see that the standard cosmology is reproduced, we rearrange the
(i,j) and $(0,0)$ gravitational field equations, assuming that the
energy-momentum tensor is comprised of matter and radiation fluids
only, obtaining

\begin{eqnarray} 3H^{2} &=& {\rho_{\rm m} + \rho_{\rm r} +
\rho_{\rm F}(H,\dot{H},\ddot{H},\dot{R}) \over M_{\rm Pl}^{2}} ,\\
2\dot{H} + 3 H^{2} &=& -{P_{\rm m} + P_{\rm r} + P_{\rm F}
(H,\dot{H},\ddot{H},\dot{R}) \over M_{\rm Pl}^{2}} ,\end{eqnarray}

\noindent where the subscripts $m$ and $r$ represent the matter
and radiation components, and $F$ denotes effective density and
pressure terms due to the $F(R)$ modified gravity function (we
have made it clear that $\rho_{\rm F}$ and $P_{\rm F}$ depend on
$H$ and its derivatives, to stress that we are dealing with a
system of fourth order differential equations). $\rho_{\rm F}$ and
$P_{\rm F}$ are given by

\begin{eqnarray}\nonumber {\rho_{\rm F} \over M_{\rm Pl}^{2}} =
&-& 3H \left[ {1 \over 3M^{2}} + {g \over \epsilon}{\rm sech}^{2}
(R/\epsilon - b)\right]\dot{R} - {R^{2} \over 12M^{2}} + (\dot{H}
+ H^{2}){R \over M^{2}} + {g\epsilon \over 2}\log(\cosh b) \\
\label{eq:w2} &-& {g \over 2} \left[\epsilon \log \cosh(R/\epsilon
- b) - R \right] + 3g(\dot{H} + H^{2})\left( \tanh ( R/\epsilon -
b) - 1\right) ,
\end{eqnarray}

\begin{eqnarray}\label{eq:w3} {P_{\rm F} \over M_{\rm Pl}^{2}} &=&
-{2c \dot{R}^{2}\over \epsilon^{2}}\tanh(R/\epsilon-b){\rm
sech}^{2} (R/\epsilon-b) \\ \nonumber &+& \left( {1 \over 3M^{2}}
+ {g \over \epsilon} {\rm sech}^{2}(R/\epsilon-b) \right)
(\ddot{R}
+ 2 H \dot{R} ) - g(\dot{H} + 3H^{2})(\tanh(R/\epsilon-b) - 1) \\
\nonumber  &-& {g \over 2} \left( R - \epsilon \log
\left[{\cosh\left(R/\epsilon - b\right) \over \cosh b}
\right]\right)  - (\dot{H} + 3H^{2}) {R \over 3 M^{2}} + {R^{2}
\over 12 M^{2}} .\end{eqnarray}

\noindent It is clear that $\rho_{\rm F}$ and $P_{\rm F}$ will act
as dark energy components, and by solving the gravitational field
equations we can obtain the equation of state parameter $w
\equiv P_{\rm F}/\rho_{\rm F}$.

To begin, we note that previous studies \cite{st} have shown that
the Ricci scalar can be written as $R = R_{\rm GR} + \delta R_{\rm
osc} + \delta R_{\rm ind}$ during the cosmological evolution of
models such as ($\ref{eq:k1}$), where $\delta R_{\rm osc}$ is the
oscillatory, scalaron component, and $\delta R_{\rm ind}$ is given
by $\delta R_{\rm ind} \simeq \Box F'(R_{\rm GR}) + R_{\rm GR}
F'(R_{\rm GR}) - 2F(R_{\rm GR}) + R_{\rm GR}$. The scalaron
oscillations $\delta R_{\rm osc}$ have been discussed in previous
sections; they are well behaved and regular for the model
($\ref{eq:k1}$), and we can assume that the energy density of
these oscillations has decayed during reheating and can be
neglected. $\delta R_{\rm ind}$ also satisfies $\delta R_{\rm ind}
\ll R_{\rm GR}$ for $R > R_{\rm vac}$, and hence as an excellent
approximation we may simply use $R \simeq R_{\rm GR}$ and $H
\simeq H_{\rm GR}$ in $\rho_{\rm F}$ and $p_{\rm F}$.

Typically during the matter and radiation era's, we have $\epsilon
\ll R_{\rm GR} \ll M^{2}$, in which case ($\ref{eq:w2}$) and
($\ref{eq:w3}$) can be written as

\begin{eqnarray}\label{eq:w5} {\rho_{\rm F} \over M_{\rm Pl}^{2}}
\simeq && {g\epsilon \over 2}\left[ b + \log (e^{b}+e^{-b})\right]
+ {1 \over  M^{2}} \left[ (\dot{H} + H^{2})R - H\dot{R} - {R^{2}
\over 12} \right] \\ \nonumber && - e^{-2(R/\epsilon - b)} \left[
{12 g H \dot{R} \over \epsilon} + {g\epsilon \over 2} + 6g(\dot{H}
+ H^{2})\right] ,\end{eqnarray}

\begin{eqnarray}\label{eq:w6} {P_{\rm F} \over M_{\rm Pl}^{2}}
\simeq && -{g \epsilon \over 2} \left[ b + \log (e^{b} +
e^{-b})\right]
 + {1 \over 12M^{2}}\left[ 4(\ddot{R} + 2H\dot{R}) + R^{2} - 4(\dot{H}
 + 3H^{2})R \right] \\ \nonumber && + e^{-2(R/\epsilon - b)}
 \left[ {g \epsilon \over 2} + 2g(\dot{H} + 3H^{2}) + {4g \over \epsilon}
 (\ddot{R} + 2H\dot{R}) - {8 g \over \epsilon^{2}}(\dot{R})^{2}\right] .\end{eqnarray}

\noindent Since last two terms in ($\ref{eq:w5}$) and
($\ref{eq:w6}$) are suppressed by factors of $1/M^{2}$ and
$e^{-2(R/\epsilon - b)}$ respectively, they are completely
subdominant, and throughout the matter and radiation era's we have
$w \approx -1$. This simply reflects the fact that for $R
\gg R_{\rm vac}$, the mass of the scalaron $M_{\rm scal}$ is very
large, and as an effective field theory below energy scales $E
\sim M_{\rm scal}$ the model ($\ref{eq:k1}$) reduces to GR with a
cosmological constant.

However, at late times, when $R_{\rm GR} \sim {\cal O}(R_{\rm
vac})$, the mass of the scalaron is small, and we expect that
there may be significant deviations from $w = -1$. The
change in $w$ is due to the $\tanh(R/\epsilon - b) -1$,
${\rm sech}^{2}(R/\epsilon -b)$ and $\log[\cosh(R/\epsilon-b)]-R$
terms in ($\ref{eq:w2},\ref{eq:w3}$), which are no longer
exponentially suppressed. This behaviour is exhibited in
fig.\ref{fig:14}, where $w$ is shown as a function of
redshift. $w(z)$ was calculated by solving the full
gravitational field equations numerically, starting at $z=4$ and
evolving to $z=0$. As initial conditions, we have assumed that $H$
and its derivatives are exactly their GR counterparts at $z=4$,
and $\Omega_{\rm m} = 0.3$, and have taken $\delta = 10^{-7}$.
With this choice, we observe that $w$ oscillates around
$-1$ in the past as expected, but at redshifts $z \simeq 1$,
$w$ drifts from its GR value and presents phantom behaviour,
$w<-1$. We note that the size of this `drift' depends
almost entirely on our choice of $b$; the larger we take $b$, the
smaller $R/\epsilon$ is and the more suppressed the
$\tanh(R/\epsilon-b)$ and ${\rm sech}^{2}(R/\epsilon-b)$ terms
become. This is clearly shown in fig.\ref{fig:14}; as we increase
$b$, the late time drift from $w=-1$ becomes increasingly
suppressed; $|\triangle w| \simeq 0.06$ for $b=1.2$ and
$|\triangle w| \simeq 2\times 10^{-3}$ for $b=4$.

Finally, we note that $w$ crosses the phantom boundary and
satisfies $w>-1$ for $z<1$ and moderate values of $b$. Thus, our model
naturally exhibits phantom behaviour at recent redshifts during the
matter dominated stage. Deviations of $w(z)$ from $-1$ are small, less
than several percent, hence there is a good agreement with present
observational upper bounds on $|w_{\rm DE}+1|$, see e.g. \cite{K08}.

\FIGURE{\hspace{10mm}\epsfig{file=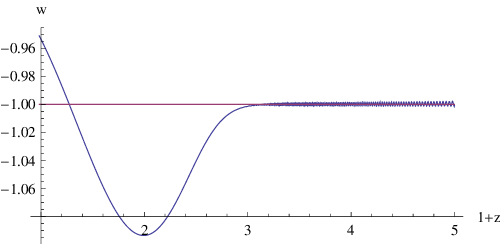,width=0.8\textwidth}\\
        \epsfig{file=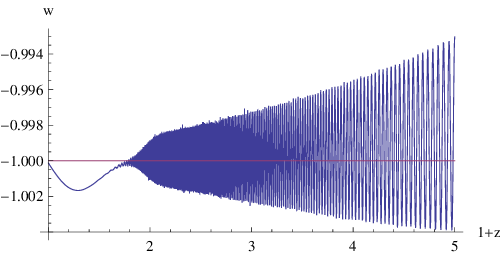,width=0.8\textwidth}
        \caption[]{\label{fig:14}The `dark energy' equation of state
parameter $w$ as a function of redshift for the $gR^2$-AB
model, with $b = 1.2$ (top) and $b=4$ (bottom). We note that
$w$ oscillates around $w=-1$ at earlier times, as
can clearly be seen in the $b=4$ case, however for redshifts $z
\sim 0$ the oscillations are small but the `drift' terms in
$\rho$ and $P_{\rm F}$ become important. We see the
deviation at $z \sim 0$ depends on $b$; as $b$ increases,
deviations from GR become increasingly suppressed.}

    }

\section{\label{sec:5}Conclusions and discussion}

To summarize, in this paper we have shown for the first time that
it is possible to construct at least one self-consistent model of
present DE in the scope of $F(R)$ gravity which satisfies all five
viability conditions presented in the Introduction, is free of new
singularities and does not destroy any of the previous successes of
cosmology. To achieve this aim, it has been necessary to extend
the range of $R$ over which previous DE $F(R)$ models were defined
to both large positive and negative values, and change the
behaviour of the models correspondingly; see Eq.\,(\ref{eq:k1})
for the improved $gR^2$-AB model.

Furthermore, since the large-$R$ behaviour (\ref{R-corr}) of the
$F(R)$ function, which is needed to avoid new singularities and to
solve two other problems of previous models, appears to be just
the same as needed for scalaron-driven inflation in $F(R)$
gravity, we have shown that the model Eq.\,(\ref{eq:k1}) can
describe inflation (primordial DE), the present acceleration of
the Universe (present DE) and the intermediate epochs of radiation
and matter domination for the unique choice of its parameter $M$,
determined by the observed power of scalar (density)
perturbations. Unexpectedly, we have found that the low-curvature
modification of $F(R)$ from its GR value, which is needed to
describe present DE, strongly affects processes at very high
values of curvature, specifically during reheating after inflation, through its
non-zero $g$-factor (\ref{g-factor}). As a result, in contrast to
pure inflationary models of $F(R)$ gravity which have $g=0$,
scalaron oscillations after the end of inflation become strongly
non-linear and the Universe evolution passes through the sequence
of interchanging periods with $a(t)\propto \sqrt t$ and $a\approx
const$, with the number of time e-folds $\Delta \ln t$ equal to
$2\ln(1/(1-2g))$ and $\ln(1/(1-2g))$ for them correspondingly. We find that on average, the Universe expands as $a(t)\propto t^{1/3}$ during this period;
this is the most significant mathematical result of the paper.

Creation of particles and antiparticles of usual matter and final
reheating are achieved by taking into account the process of
gravitational particle creation by these non-linear oscillations
of $R$. This mainly occurs at the end of inflation, so reheating is
less efficient than in the purely inflationary $R+R^2/6M^2$ model,
although still viable. Due to the different average law of
expansion after the end of scalaron-driven inflation, predictions
for parameters of primordial spectra of scalar (density)
perturbations and gravitational waves generated during inflation
in this combined model of primordial and present DE are slightly
different to those for the inflationary model only.
The difference is due to the change in the number of scale factor
e-folds $N$ used in the corresponding formulas: from $N\approx
(50-55)$ to $N=70$. This specific prediction of the combined model
is observable, however, it is degenerate with a possible slow
variation of the $R^2$ behaviour of $F(R)$ at large $R$ in the
model (\ref{eq:k1}), e.g. like in Eq.\,(\ref{chaotic}).

In this paper we have not addressed the issue of neutron star
stability in $F(R)$ gravity raised in \cite{afq1} and further
considered in \cite{ma,KM09}, since the most recent results in
\cite{la1,huu1} suggest that there is no problem and
that, for our choice of the parameter $M$ in the improved model
(\ref{eq:k1}), non-GR corrections are very small if the trace $T$
of the matter energy-momentum tensor remains non-positive inside
neutron stars. However, due to a non-zero $g$-factor, this problem
becomes highly non-trivial and requires special consideration if
$T$ is permitted to become positive at large matter densities.

The most critical observational prediction of DE models in
$F(R)$ gravity remains the anomalous growth of density
perturbations in the matter component at recent redshifts
\cite{st,T07,SPH07,GMP09,MSY09} which results, in particular, in a
mismatch between parameters such as
$\sigma_8$ and $n_s$ determined from CMB temperature anisotropy
and galaxy clustering separately, assuming GR (as well as from the
cluster abundance at different $z$). The absence of such an effect at
the level of $\sim 5\%$ in the HSS model makes its background
evolution practically indistinguishable from that in the standard
$\Lambda$CDM. However, this effect is suppressed in the AB model
as compared to the HSS one, so it is more difficult to falsify the
former model. Future observational data will determine the fate of
this whole class of models.

\section*{Acknowledgements}

AAS was partially supported by the grant RFBR 08-02-00923 and by
the Scientific Programme ``Astronomy'' of the Russian Academy of
Sciences.

\end{document}